\documentclass[10pt,journal,compsoc]{IEEEtran}
\usepackage{graphicx, amsmath}
\usepackage{array}
\usepackage{multirow}
\usepackage{booktabs}
\usepackage{makecell}
\usepackage[table,xcdraw]{xcolor}

  \usepackage{cite}
  \usepackage[hyphens]{url}
\ifCLASSINFOpdf
\else
\fi
\begin{document}
%
\title{Audio Deepfake Detection: A Survey}
%
%
%
%

\author{Jiangyan~Yi,~\IEEEmembership{Member,~IEEE,}
        Chenglong~Wang,
        Jianhua~Tao,~\IEEEmembership{Senior~Member,~IEEE,}
        Xiaohui~Zhang,
        Chu~Yuan~Zhang,
        and Yan Zhao
\IEEEcompsocitemizethanks{\IEEEcompsocthanksitem Jiangyan Yi and Chu Yuan Zhang are
with the State Key Laboratory of Multimodal Artificial Intelligence Systems, Institute of Automation, Chinese Academy of Sciences and
University of Chinese Academy of Sciences. E-mail: jiangyan.yi@nlpr.ia.ac.cn (Jiangyan Yi and Jianhua Tao are corresponding authors.)
\IEEEcompsocthanksitem Jianhua Tao is with the Department of Automation, Tsinghua University.
\IEEEcompsocthanksitem Chenglong Wang is with the University of Science and Technology of China.
\IEEEcompsocthanksitem Xiaohui Zhang is with the Beijing Jiaotong University.
\IEEEcompsocthanksitem Yan Zhao is with the Hebei University of Technology.
\protect\\
}
}

%
%

\markboth{Journal of \LaTeX\ Class Files,~Vol.~14, No.~8, August~2023}%
{Shell \MakeLowercase{\textit{et al.}}: Bare Demo of IEEEtran.cls for Computer Society Journals}
%



\IEEEtitleabstractindextext{%
\begin{abstract}

Audio deepfake detection is an emerging active topic. A growing number of literatures have aimed to study deepfake detection algorithms and achieved effective performance, the problem of which is far from being solved. Although there are some review literatures, there has been no comprehensive survey that provides researchers with a systematic overview of these developments with a unified evaluation. Accordingly, in this survey paper, we first highlight the key differences across various types of deepfake audio, then outline and analyse competitions, datasets, features, classifications, and evaluation of state-of-the-art approaches. For each aspect, the basic techniques, advanced developments and major challenges are discussed. In addition, we perform a unified comparison of representative features and classifiers on ASVspoof 2021, ADD 2023 and In-the-Wild datasets for audio deepfake detection, respectively. The survey shows that future research should address the lack of large scale datasets in the wild, poor generalization of existing detection methods to unknown fake attacks, as well as interpretability of detection results etc. 

\end{abstract}

}

\maketitle

\IEEEdisplaynontitleabstractindextext

%
\IEEEpeerreviewmaketitle

\IEEEraisesectionheading{\section{Introduction}\label{sec:introduction}}

%
%
%
%
\IEEEPARstart{O}{ver} the past few years, deep learning based text-to-speech (TTS) and voice conversion (VC) technologies have made great improvement~\cite{Wang2021Prosody, Yi2020VCC}. These technologies enable the generation of human-like natural speech that proves difficult to distinguish from real audio. Admittedly, the development in these technologies significantly improve the conveniences of our life in many scenarios, such as in-car navigation systems, e-book readers, intelligent robots etc. They nonetheless also pose a serious threat to social security and political economy if someone misuses the so-called deepfake technologies for malicious purposes. The term~\textit{deepfake}~\cite{B2019Deepfake} refers to the usage of deep learning methods to seamlessly swap faces in videos on Reddit in 2017. Nowadays, ~\textit{deepfake} is now generically used by the media or people to refer to any audio or video in which important attributes have been either digitally altered or swapped, with the help of artificial intelligence (AI). Fraudsters used AI based software to mimic a chief executive’s voice and demand a fraudulent transfer of USD 243, 000 in 2019~\cite{Stupp2019mimicvoice}.
In response to these attacks, it is necessary to be able to detect deepfake audio.

\begin{figure}[htb]
\hfill
\begin{minipage}[b]{1.0\linewidth}
  \centering
  \centerline{\includegraphics[width=9.0cm,height=2.6cm]{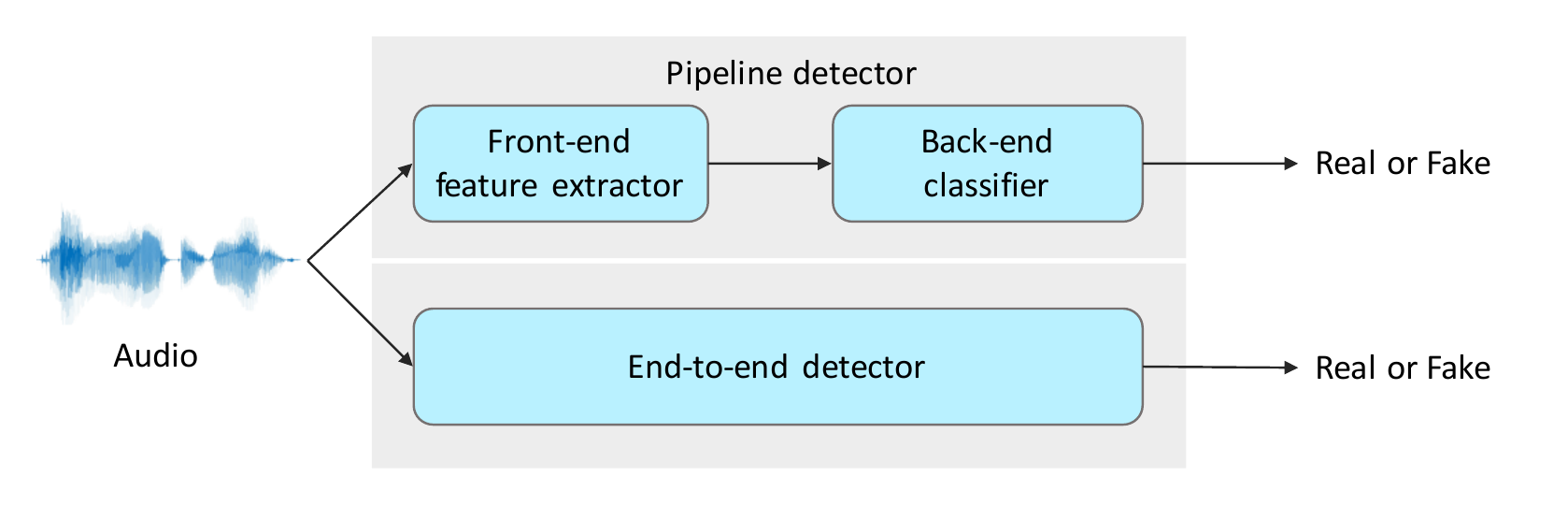}}
\end{minipage}
\caption{Mainstream solutions on audio deepfake detection: pipeline and end-to-end detector.}
\label{fig:detection}
\end{figure}

Audio deepfake detection is a task that aims to distinguish genuine utterances from fake ones via machine learning techniques as shown in Figure~\ref{fig:detection}. An increasing number of attempts~\cite{Chen2020Gen, Wang2020Deep, Wu2020LCNN} have been made to further the development of audio deepfake detection.
Existing mainstream studies on audio deepfake detection can be roughly categorized into two kinds of solutions: pipeline and end-to-end detector. The pipeline solution, consisting of a front-end feature extractor and a back-end classifier, has become the de facto standard framework over the last decades. In recent years, end-to-end methods have attracted more and more attention, which employ a model to jointly optimise the feature extraction and classification via operating directly upon raw audio waveform.

\begin{figure*}[htb]
\hfill
\begin{minipage}[b]{0.31\linewidth}
  \centering
  \centerline{\includegraphics[width=5.2cm,height=2.2cm]{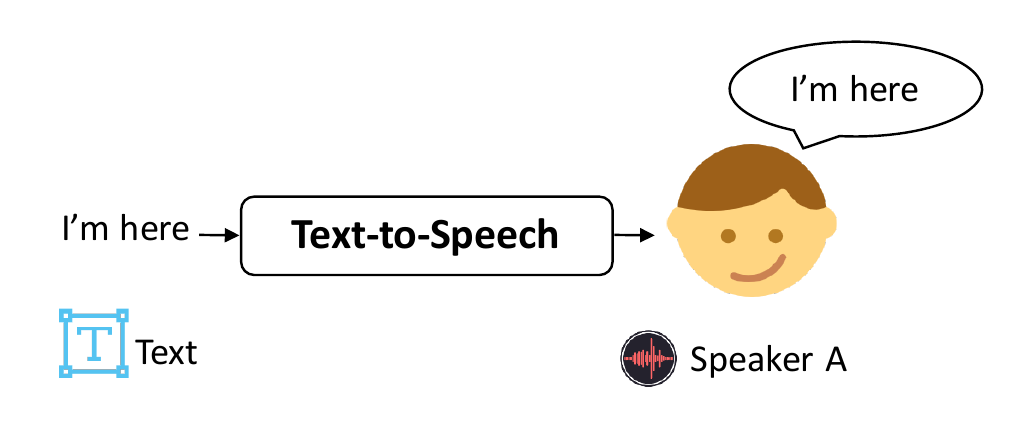}}
  \centerline{(a)}\medskip
\end{minipage}
\hfill
\begin{minipage}[b]{0.32\linewidth}
  \centering
  \centerline{\includegraphics[width=5.5cm,height=2.3cm]{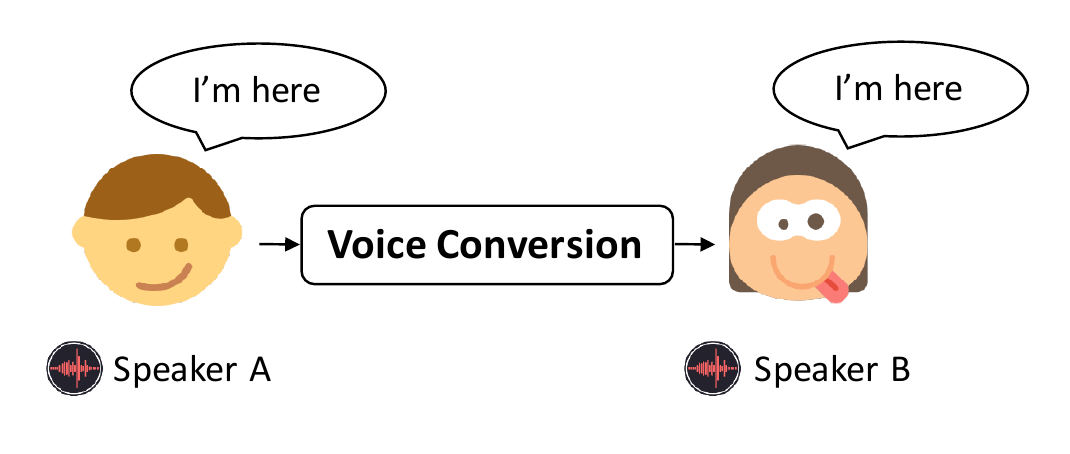}}
  \centerline{(b)}\medskip
\end{minipage}
\hfill
\begin{minipage}[b]{0.34\linewidth}
  \centering
  \centerline{\includegraphics[width=5.7cm,height=2.4cm]{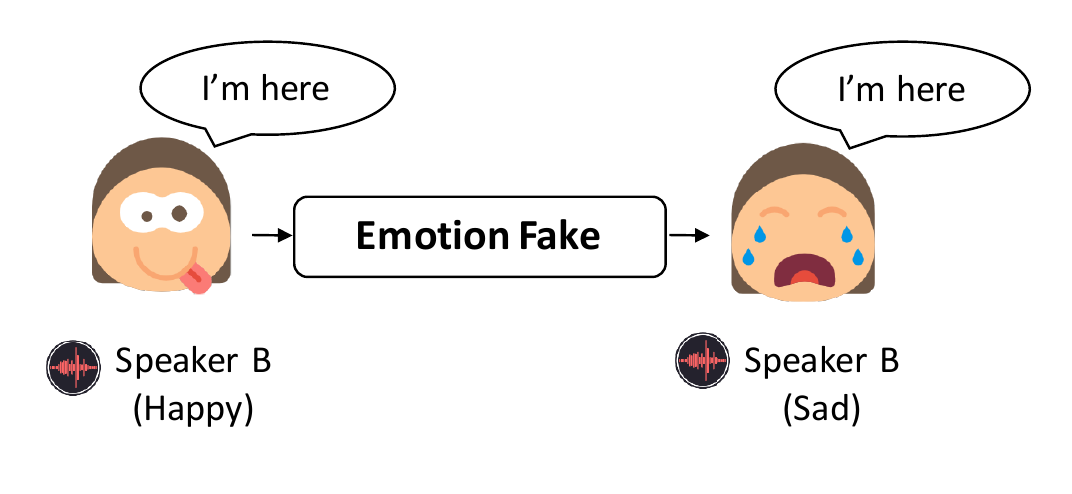}}
  \centerline{(c)}\medskip
\end{minipage}
\hfill
\begin{minipage}[b]{0.48\linewidth}
  \centering
  \centerline{\includegraphics[width=6.5cm,height=2.5cm]{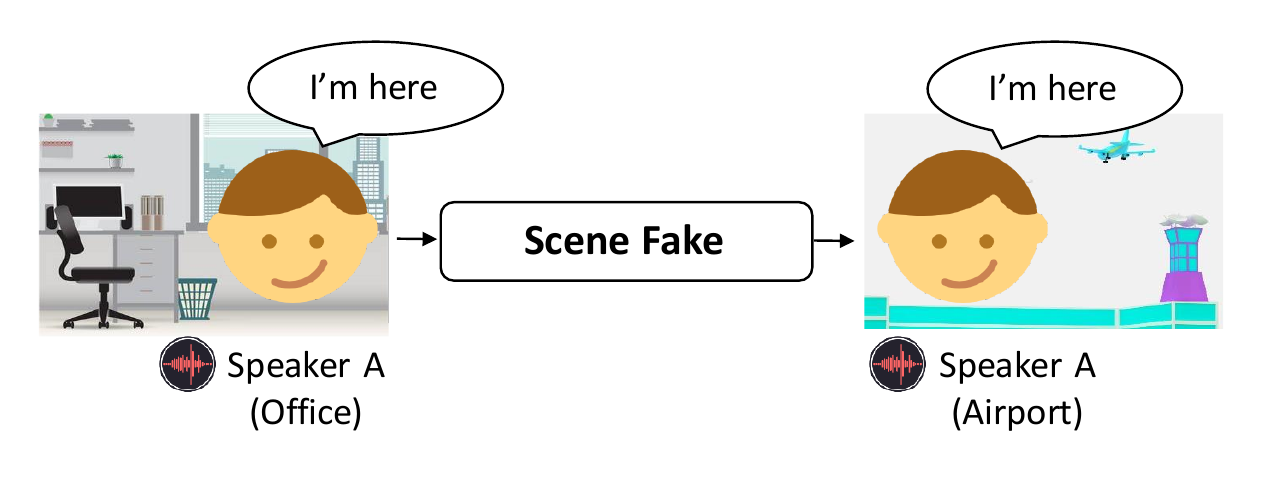}}
  \centerline{(d)}\medskip
\end{minipage}
\hfill
\begin{minipage}[b]{0.49\linewidth}
  \centering
  \centerline{\includegraphics[width=8.0cm,height=2.4cm]{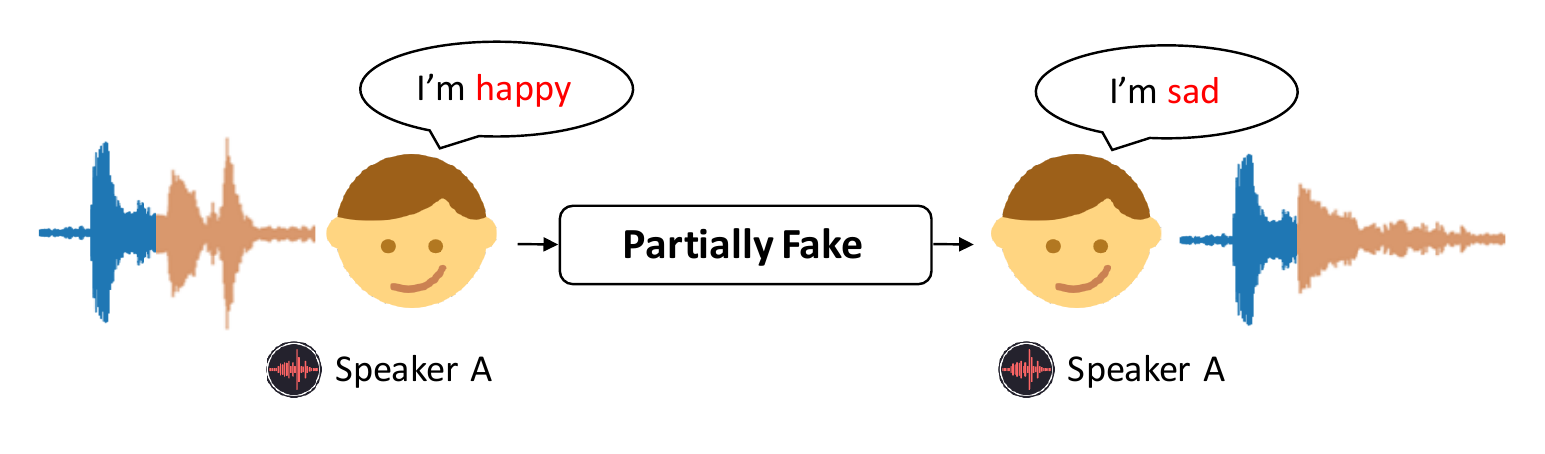}}
  \centerline{(e)}\medskip
\end{minipage}
\caption{Five kinds of deepfake audio: (a) text-to-speech, (b) voice conversion, (c) emotion fake, (d) scene fake, (e) partially fake.}
\label{fig:faketypes}
\end{figure*}

Although previous studies on audio deepfake detection have obtained promising performance, their scopes remain largely scattered, with few systematic surveys. Most of them aim for summarising previous spoofing attacks and countermeasures for protecting automatic speaker verification (ASV) systems. Wu et al.~\cite{Wu2015Spoofing} provide a comprehensive survey of past work to assess the vulnerability of ASV systems and the countermeasures to protect them in 2015. One recent review literature~\cite{2020Advances} presents advances in anti-spoofing from a perspective of ASVspoof challenges in 2020. Another survey work~\cite{2021survey} presents and analyses attack detection work for ASV systems published between 2015 and 2021. Aakshi et al.~\cite{Mittal2021ASV} review and analysis most of the benchmark spoofed speech datasets, methods and evaluation metrics for ASV systems and spoof detection techniques. Very few of them focus on summarising the past work of audio deepfake detection through the lens of helping people refrain from being deceived. Most recently, a survey~\cite{2022Review} introduces the deepfake audio types, datasets and detection methods.
But it only provides a collection of results from classic methods, and lacks consistent experimental analysis.

Different from~\cite{2022Review}, this paper presents a comprehensive survey that makes the following contributions. We provide a systematic overview focusing on learning common discriminative audio features related to audio deepfake detection, as well as computing methodologies that can be used to build an appropriate generalized automatic system. This survey also includes a detailed summary of up-to-date audio deepfake detection datasets; based on this summary, we perform a unified comparison of representative detection methods.

The remainder of this paper is organized as follows. Section 2 highlights differences across various types of deepfake audio, summarizes existing benchmark datasets and competitions, as well as evaluation metrics. Discriminative features for audio deepfake detection are presented and categorized in Section 3. Representative classification algorithms are summarized in Sections 4. End-to-end methods and generalization methods are introduced in Sections 5 and 6. Section 7 presents a detailed comparison of different features and models. Some remaining challenges and future research directions are summarized in Section 8. Finally, Section 9 concludes this paper.

\section{Overview}
The field of audio deepfake detection has been blossoming in terms of deepfake technologies, competitions, datasets, evaluation metrics and detection methods. 

\subsection{Types of Deepfake Audio}
Deepfake audio generally refers to any audio in which important attributes have been manipulated via AI technologies while still retaining its perceived naturalness. Previous studies mainly involve five kinds of deepfake audio: text-to-speech, voice conversion, emotion fake, scene fake, partially fake. The characteristics of different deepfake types are summarised in Table \ref{table:faketypes}.

\begin{table}[]
  \caption{Summary of audio deepfake types in past studies.}
  \label{table:faketypes}
  \renewcommand
  \arraystretch{1.5}
  \centering
  \scriptsize  
  \begin{tabular}{cccc}
  \toprule[1pt]
  \textbf{Fake Type} & \textbf{Fake Trait}  & \textbf{Fake Duration} & \textbf{AI-aided} \\\hline
  Text-to-speech (TTS) & \makecell[c]{Speaker identity,\\Speech content}  & Fully & Yes \\
  \rowcolor[HTML]{EFEFEF} 
  Voice conversion (VC) & Speaker identity  & Fully & Yes \\
  Emotion fake & Speaker emotion & Fully & Yes \\
  \rowcolor[HTML]{EFEFEF} 
  Scene fake & Acoustic scene  & Fully & Yes \\
  Partially fake & Speech content & Partially & Yes \\
  \bottomrule[1pt] 
  \end{tabular} 
\end{table}

\begin{table*}[]
  \caption{Characteristics of representative competitions on audio fake detection.}
  \label{table:competitions}
	\resizebox{\textwidth}{!}{
	\begin{tabular}{ccccccccccc}
		\toprule[1pt]
		\textbf{Competition} & \textbf{Language} & \textbf{Year} & \textbf{\#Registration} & \textbf{\#Submission} & \textbf{Task} & \textbf{Fake Type} & \textbf{Deepfake} & \textbf{Goal} &  \multicolumn{2}{c}{\textbf{Baseline Model}} \\
		\cmidrule(r){10-11} 
		& & & & & & & & & \textbf{Features} & \textbf{Classifiers} \\
		\hline
		&  & \cellcolor[HTML]{EFEFEF}2015 & \cellcolor[HTML]{EFEFEF}28 & \cellcolor[HTML]{EFEFEF}16 & \cellcolor[HTML]{EFEFEF}LA & \cellcolor[HTML]{EFEFEF}TTS, VC & \cellcolor[HTML]{EFEFEF}No & \cellcolor[HTML]{EFEFEF}Detection & \cellcolor[HTML]{EFEFEF}- & \cellcolor[HTML]{EFEFEF}- \\
		&  & \cellcolor[HTML]{FFFFFF}2017 & \cellcolor[HTML]{FFFFFF}113 & \cellcolor[HTML]{FFFFFF}49 & \cellcolor[HTML]{FFFFFF}PA & \cellcolor[HTML]{FFFFFF}Replay & \cellcolor[HTML]{FFFFFF}No & \cellcolor[HTML]{FFFFFF}Detection & \cellcolor[HTML]{FFFFFF}CQCC& \cellcolor[HTML]{FFFFFF}GMM \\
		&  & \cellcolor[HTML]{EFEFEF} & \cellcolor[HTML]{EFEFEF} & \cellcolor[HTML]{EFEFEF}48 & \cellcolor[HTML]{EFEFEF}LA & \cellcolor[HTML]{EFEFEF}TTS, VC & \cellcolor[HTML]{EFEFEF}No & \cellcolor[HTML]{EFEFEF}Detection &\cellcolor[HTML]{EFEFEF} & \cellcolor[HTML]{EFEFEF}\\
		&  & \multirow{-2}{*}{\cellcolor[HTML]{EFEFEF}2019} & \multirow{-2}{*}{\cellcolor[HTML]{EFEFEF}154} & \cellcolor[HTML]{EFEFEF}50 & \cellcolor[HTML]{EFEFEF}PA & \cellcolor[HTML]{EFEFEF}Replay & \cellcolor[HTML]{EFEFEF}No & \cellcolor[HTML]{EFEFEF}Detection & \cellcolor[HTML]{EFEFEF}\multirow{-2}{*}{LFCC, CQCC}& \cellcolor[HTML]{EFEFEF}\multirow{-2}{*}{GMM} \\
		&  & \cellcolor[HTML]{FFFFFF} & \cellcolor[HTML]{FFFFFF} & \cellcolor[HTML]{FFFFFF}41 & \cellcolor[HTML]{FFFFFF}LA & \cellcolor[HTML]{FFFFFF}TTS, VC & \cellcolor[HTML]{FFFFFF}No & \cellcolor[HTML]{FFFFFF}Detection & \cellcolor[HTML]{FFFFFF} & \cellcolor[HTML]{FFFFFF}\\
		&  & \cellcolor[HTML]{FFFFFF} & \cellcolor[HTML]{FFFFFF} & \cellcolor[HTML]{FFFFFF}23 & \cellcolor[HTML]{FFFFFF}PA & \cellcolor[HTML]{FFFFFF}Replay & \cellcolor[HTML]{FFFFFF}No & \cellcolor[HTML]{FFFFFF}Detection & \cellcolor[HTML]{FFFFFF} & \cellcolor[HTML]{FFFFFF} \\
		\multirow{-7}{*}{ASVspoof} & \multirow{-7}{*}{English} & \multirow{-3}{*}{\cellcolor[HTML]{FFFFFF}2021} & \multirow{-3}{*}{\cellcolor[HTML]{FFFFFF}198} & \cellcolor[HTML]{FFFFFF}33 & \cellcolor[HTML]{FFFFFF}DF & \cellcolor[HTML]{FFFFFF}Deepfake & \cellcolor[HTML]{FFFFFF}Yes & \cellcolor[HTML]{FFFFFF}Detection & \cellcolor[HTML]{FFFFFF}\multirow{-3}{*}{LFCC, CQCC, Raw} & \cellcolor[HTML]{FFFFFF}\multirow{-3}{*}{GMM, LCNN,  RawNet2} \\ \hline
		&  & \cellcolor[HTML]{EFEFEF} & \cellcolor[HTML]{EFEFEF} & \cellcolor[HTML]{EFEFEF}48 & \cellcolor[HTML]{EFEFEF}LF & \cellcolor[HTML]{EFEFEF}TTS, VC & \cellcolor[HTML]{EFEFEF}Yes & \cellcolor[HTML]{EFEFEF}Detection & \cellcolor[HTML]{EFEFEF} & \cellcolor[HTML]{EFEFEF} \\
		&  & \cellcolor[HTML]{EFEFEF} & \cellcolor[HTML]{EFEFEF} & \cellcolor[HTML]{EFEFEF}33 & \cellcolor[HTML]{EFEFEF}PF & \cellcolor[HTML]{EFEFEF}Partially fake & \cellcolor[HTML]{EFEFEF}Yes & \cellcolor[HTML]{EFEFEF}Detection & \cellcolor[HTML]{EFEFEF} & \cellcolor[HTML]{EFEFEF} \\
		&  & \multirow{-3}{*}{\cellcolor[HTML]{EFEFEF}2022} & \multirow{-3}{*}{\cellcolor[HTML]{EFEFEF}121} & \cellcolor[HTML]{EFEFEF}39 & \cellcolor[HTML]{EFEFEF}FG & \cellcolor[HTML]{EFEFEF}TTS, VC & \cellcolor[HTML]{EFEFEF}Yes & \cellcolor[HTML]{EFEFEF}Game fake & \cellcolor[HTML]{EFEFEF}\multirow{-3}{*}{LFCC, Raw} & \cellcolor[HTML]{EFEFEF}\multirow{-3}{*}{GMM, LCNN, RawNet2} \\
		&  & \cellcolor[HTML]{FFFFFF} & \cellcolor[HTML]{FFFFFF} & \cellcolor[HTML]{FFFFFF}63 & \cellcolor[HTML]{FFFFFF}FG & \cellcolor[HTML]{FFFFFF}TTS, VC & \cellcolor[HTML]{FFFFFF}Yes & \cellcolor[HTML]{FFFFFF}Game fake & \cellcolor[HTML]{FFFFFF}LFCC, Wav2vec2.0 & GMM, LCNN \\
		&  & \cellcolor[HTML]{FFFFFF} & \cellcolor[HTML]{FFFFFF} & \cellcolor[HTML]{FFFFFF}16 & \cellcolor[HTML]{FFFFFF}RL & \cellcolor[HTML]{FFFFFF}Partially fake & \cellcolor[HTML]{FFFFFF}Yes & \cellcolor[HTML]{FFFFFF}Forensics & \cellcolor[HTML]{FFFFFF}LFCC & LCNN \\
		\multirow{-6}{*}{ADD} & \multirow{-6}{*}{Chinese} & \multirow{-3}{*}{\cellcolor[HTML]{FFFFFF}2023} & \multirow{-3}{*}{\cellcolor[HTML]{FFFFFF}145} & \cellcolor[HTML]{FFFFFF}11 & \cellcolor[HTML]{FFFFFF}AR & \cellcolor[HTML]{FFFFFF}TTS, VC & \cellcolor[HTML]{FFFFFF}Yes & \cellcolor[HTML]{FFFFFF}Attribution & \cellcolor[HTML]{FFFFFF}LFCC & ResNet + Openmax \\ \bottomrule[1pt]
	\end{tabular}
}
\end{table*}

\subsubsection{Text-to-Speech}
Text-to-speech (TTS)~\cite{Wu2015Spoofing}, commonly known as speech synthesis and shown in Figure 2 (a), aims to synthesise intelligible and natural speech, given any arbitrary text, using machine learning based models. TTS models can generate realistic and human-like speech with the development of deep neural networks~\cite{Wang2021Prosody}. TTS systems mainly include text analysis and speech waveform generation modules. There are two major methods on speech waveform generation: concatenative~\cite{1996Unit,2001Applying} and statistical parametric TTS~\cite{Zen2013StaTTS}. The latter often consists of an acoustic model and a vocoder. Most recently, some end-to-end models have been proposed to generate high-quality sounding audio, such as Variational Inference with adversarial learning for end-to-end Text-to-Speech (VITS)~\cite{Kim2021VITS} and FastDiff-TTS~\cite{Huang2022FastDiff}. 

\subsubsection{Voice Conversion}
Voice conversion (VC)~\cite{Wu2015Spoofing} refers to cloning a person's voice digitally as shown in Figure~\ref{fig:faketypes}~(b). It aims to change the timbre and prosody of a given speaker's speech to that of another speaker, while the content of the speech remians the same. The input to a VC system is a natural utterance of the given  speaker. There are about three main approaches of VC technologies: statistical parametric~\cite{sisman2020or, choi2021neural}, frequency warping~\cite{ godoy2011voice} and unit-selection~\cite{jin2016cute}. Statistical parametric model also has a vocoder which is similiar to that in statistical parametric TTS~\cite{oord2016wavenet, kong2020hifi}. In recent years, end-to-end VC models have also been proposed to mimic a person's voice characteristics~\cite{dahmani2019vits}.

\subsubsection{Emotion Fake}
Emotion fake~\cite{Zhao2022EmoFake} seeks to change the audio in such a way that the emotion of the speech changes, while other information remains the same, such as speaker identity and speech content. Changing emotions of the voice often leads to semantics changes. An example of emotion fake is illustrated in Figure~\ref{fig:faketypes}~(c). The original utterance said by speaker B is with a happy emotion. The fake utterance is the audio where the happy emotion has been changed into a sad emotion. There are two kinds of methods on emotional VC called emotion fake~\cite{Zhou2021EmoVC}: parallel data based and non-parallel data based methods.

\subsubsection{Scene Fake}
Scene fake~\cite{Yi2022SceneFake} involves the tempering of the acoustic scene of the original utterance with another scene via speech enhancement technologies while the speaker identity and speech content remain unchanged. An example of scene fake is shown in Figure~\ref{fig:faketypes}~(d). The acoustic scene of the real utterance is~``\textit{Office}''. The acoustic scene of the fake utterance is~``\textit{Airport}''. If the scene of an original audio is manipulated with another one, authenticity and integrity verification of the audio will be unreliable and even the semantic meaning of the original audio could be changed.

\subsubsection{Partially Fake}
Partially fake~\cite{Yi2021Half} focuses on only changing several words in an utterance. The fake utterance is generated by manipulating the original utterances with genuine or synthesized audio clips. The speaker of the original utterance and fake clips is the same person. The synthesized audio clips, while keeping the speaker identity unchanged. An example of partially fake is shown in Figure~\ref{fig:faketypes}~(e). 

\subsection{Competitions}
Over the last few years, a series of competitions have played a key role in accelerating the development of audio deepfake detection, such as the ASVspoof\footnote{\url{https://www.asvspoof.org}} and ADD\footnote{\url{http://addchallenge.cn}} challenges. Table~\ref{table:competitions} shows the characteristics and baseline models of the representative competitions.

The ASVspoof challenges mainly focus on detecting spoofed audio from the perspective of protecting ASV systems from attack. The ASVspoof 2015~\cite{Wu2015ASVspoof} involves logical access (LA) task involving the detection of synthetic and converted utterances. The ASVspoof 2017~\cite{2017ASVspoof} only has one task named physical access (PA), which includes replay attacks. The ASVspoof 2019~\cite{2019ASVspoof} consists of two tasks: LA and PA, which are included in previous two challenges. Speech deepfake detection task is included in the ASVspoof 2021~\cite{2021ASVspoof}, which consists of three tasks: LA, PA and speech deepfake (DF). The DF task involves compressed audio similar to the LA task.

The ADD 2022 challenge~\cite{Yi2022ADD} was organized by including three tasks: low-quality fake audio detection (LF), partially fake audio detection (PF) and audio fake game (FG). The LF task focuses on dealing with genuine and fully fake utterances with various real-world noises and interferences etc. The PF task aims to distinguish between partially fake and real audio. The FG task is a rivalry game involving an audio generation task and an audio fake detection task, wherein the generation task participants aim to generate audio that could deceive the detection systems submitted by the detection task participants. The results in ADD 2022 show that it is difficult to use the same model to deal with all fake types. The result also show that the generalisation of detection techniques remains an open problem. Different from previous challenges (e.g. ADD 2022), ADD 2023~\cite{Yi2023ADD} focuses on surpassing the constraints of binary real or fake classification, and actually localizing the manipulated intervals in a partially fake utterance as well as pinpointing the source responsible for generating any fake audio. The ADD 2023 challenge includes three subchallenges: audio fake game (FG), manipulation region location (RL) and deepfake algorithm recognition (AR). 

\begin{table*}[]
  \caption{Characteristics of representative datasets on audio deepfake detection. SR denotes sampling rate and SL refers to average length of utterances. Utt and Spk denote utterances and speakers, respectively.}
  \label{table:datasets}
  \centering
  \resizebox{\textwidth}{!}{
  \begin{tabular}{ccccccccccc}
  	\toprule[1pt]
   & \textbf{ASVspoof 2021} & \multicolumn{3}{c}{\textbf{ADD 2022}} & \multicolumn{3}{c}{\textbf{ADD 2023}} & \textbf{In-the-Wild} & \textbf{WaveFake} & \textbf{FoR}  \\
   \cmidrule(r){3-5} \cmidrule(r){6-8}
   & \textbf{DF} & \textbf{LF} & \textbf{PF} & \textbf{FG-D} & \textbf{FG-D} & \textbf{LR} & \textbf{AR} & {\color[HTML]{FF0000} \textbf{}} & \textbf{} & \textbf{}  \\
   \hline
   \rowcolor[HTML]{EFEFEF}
  Year & 2021 & 2022 & 2022 & 2022 & 2023 & 2023 & 2023 & 2022 & 2021 & 2019 \\
  Language & English & Chinese & Chinese & Chinese & Chinese& Chinese & Chinese & English & English & English  \\
  \rowcolor[HTML]{EFEFEF}
  Goal & Detection & Detection & Detection & Game fake & Game fake & Forensics & Attribution & Detection & Detection & Detection  \\
  Fake Types & \begin{tabular}[c]{@{}c@{}} VC, TTS\end{tabular} & TTS, VC & Partially fake & \begin{tabular}[c]{@{}c@{}}TTS, VC\end{tabular} & TTS, VC & Partially fake & TTS, VC & TTS & TTS & TTS  \\
  \rowcolor[HTML]{EFEFEF}
  Condition & \begin{tabular}[c]{@{}c@{}}Clean,\\ Noisy\end{tabular} & Noisy & Clean &  \begin{tabular}[c]{@{}c@{}}Clean,\\ Noisy\end{tabular} & \begin{tabular}[c]{@{}c@{}}Clean,\\ Noisy\end{tabular}& Noisy &Clean &Noisy  & Clean & Clean  \\
  Format & FLAC & WAV & WAV & WAV & WAV & WAV & WAV & WAV & WAV & WAV \\
  \rowcolor[HTML]{EFEFEF}
  SR (Hz) & 16k & 16k & 16k & 16k & 16k & 16k & 16k & 16k & 16k & 16k  \\
  SL (s) & 0.5$\sim$12 & 1$\sim$10 & 1$\sim$10  & 1$\sim$10 & 1$\sim$10 & 1$\sim$10 & 1$\sim$10 & 2$\sim$8 & 8$\sim$12  & 0.5$\sim$20  \\
  \rowcolor[HTML]{EFEFEF}
  \#Hours & 325.8 & 222.0 & 201.8 & 396.0 &394.7  &  131.2 & 194.5 & 38.0 & 196.0 &   150.3 \\
  \#Real Utt & 22, 617 & 36, 953 & 23, 897 & 46, 871  & 172, 819 & 46, 554 & 14, 907 & 19, 963 & 0 & 108, 256  \\
  \rowcolor[HTML]{EFEFEF}
  \#Fake Utt & 589, 212  & 123, 932 & 127, 414 & 243, 537 & 113, 042 & 65, 449 & 95, 383 & 11, 816 & 117, 985 & 87, 285  \\
  \#Real Spk & 48 & \textgreater 400  & \textgreater 200 & \textgreater 400  & \textgreater 1000  & \textgreater 200 & \textgreater 500 & 58 & 0 & 140  \\
  \rowcolor[HTML]{EFEFEF}
  \#Fake Spk & 48 & \textgreater 300 & \textgreater 200 & \textgreater 300 & \textgreater 500 & \textgreater 200 & \textgreater 500 & 58 & 2 & 33 \\
  Accessibility & Public & Restricted & Restricted & Restricted  & Restricted & Restricted & Restricted &Public  & Public & Public \\
  \bottomrule[1pt]
  \end{tabular}
  }
\end{table*}

\subsection{Benchmark Datasets}
The development of audio deepfake detection techniques has been largely dependent on well-established datasets with various fake types and diverse acoustic conditions. A variety of datasets have been designed to protect ASV systems or human listeners from spoofing or deceiving.  Table~\ref{table:datasets} highlights the characteristics of representative datasets on audio deepfake detection.

Many early studies designed spoofed datasets to develop spoofing countermeasures for ASV systems.
In the early days, a diverse set of spoofing datasets were proprietary due to the design of a dataset depending very much on the specific spoofing approach assumed in a particular study.
Some spoofing datasets are designed only involving a kind of TTS method\cite{Leon2010Eval} or a sort of VC approach~\cite{Alegre2013A, Kons2013Voice}. However, it is difficult to make comparisons across different spoofing methods. To alleviate this issue, several spoofed datasets including multiple approaches are designed by Wu et al.~\cite{Wu2013Vulnerability} and Alegre et al.~\cite{Alegre2013A}, which involve replay, TTS and VC technologies. But the varieties of spoofing techniques are still insufficient compared to the diversity required by generalised countermeasure studies.
In order to conduct repeatable and comparable spoofing detection studies, Wu et al.~\cite{Wu2015SAS} develop a standard public spoofing dataset SAS which consists of various TTS and VC methods in 2015. 
The SAS dataset is used to support the first ASVspoof challenge (ASVspoof 2015) aiming to detect the spoofed speech~\cite{Wu2015ASVspoof}. Replay is considered as a lowcost and challenging attack included in the ASVspoof 2017 challenge~\cite{Kinnunen2017ASVspoof}. The ASVspoof 2019 and 2021 datasets~\cite{Wang2020ASVspoof} both consist of replay, TTS and VC attacks.
Previous datasets in ASVspoof challenges focus on detecting speech attacks in the microphone channel. Lavrentyeva et al.~\cite{Galina2019Phone} design a PhoneSpoof dataset for speaker verification systems, in which the utterances are collected in telephone channels. A partially spoofed database~\cite{Zhang2021An} is designed by using voice activity detection technologies to randomly concatenate spoofed utterances.

A few audio deepfake detection datasets have been developed to protect people from deceiving by deepfake audio. The deepfake types contained in the datasets mainly include: TTS, VC, emotion fake, scene fake and partially fake.
In 2020, Reimao et al.~\cite{Reimao2020For} developed a publicly available dataset FoR containing synthetic utterances, which are generated with open-sourced TTS tools. A private fake dataset is constructed using the open-sourced VC and TTS systems~\cite{Wang2020Deep}. In 2021, Frank et al.~\cite{Frank2021WaveFake} developed a fake audio dataset named WaveFake, which contained two speakers' fake utterances synthesised by the latest TTS models. Audio deepfake attacks are included in ASVspoof 2021~\cite{2021ASVspoof}, which consider data compression effects. However, these datasets have not covered some real-life challenging situations. The datasets in ADD 2022 challenge are designed to fill the gap~\cite{Yi2022ADD}. The fake utterances in LF dataset are generated using the latest state-of-the-art TTS and VC models, which contain diversified noise interference. The fake utterances in PF dataset are chosen from the HAD dataset designed by Yi et al.~\cite{Yi2021Half}, which are generated by manipulating the original genuine utterances with real or synthesized audio segments of several key words, such as named entities. The detection task dataset of the FG track (FG-D) are randomly selected from the submitted utterances of generation task in ADD 2022. A Chinese synthetic speech detection dataset FMFCC-A~\cite{Zhang2021FMFCC} contains 13 types of fake audio involving noise addition and audio compression.
The above-mentioned datasets have played a pivotal role in accelerating the development of audio deepfake detection. However, the fake utterances mainly involve changing speaker identity, speech content or channel noise of the original audio. Most recently, Zhao et al.~\cite{Zhao2022EmoFake} design an emotion fake audio detection dataset named EmoFake, where the original emotion of a speaker's speech has been manipulated with another one but other information still remains the same. A scene manipulation audio dataset named SceneFake is constructed by Yi et al.~\cite{Yi2022SceneFake}, in which the acoustic scene of an original utterance is replaced with another one using speech enhancement technologies. In 2022, a real-world dataset named In-the-Wild are collected from publicly available sources such as social networks and popular video sharing platforms, where the utterances are from English-speaking celebrities and politicians~\cite{NM2022Does}. 

\subsection{Evaluation Metrics}

Previously, equal error rate (EER) is used as the evaluation metrics for audio deepfake detection tasks in the ASVspoof~\cite{2021ASVspoof} and ADD~\cite{Yi2022ADD} challenges. The 'threshold-free' EER is defined as follows. Let $P_{fa}(\theta)$ and $P_{miss}(\theta)$ denote the false alarm and miss rates at threshold $\theta$.

\vspace{-10pt}
\begin{align}
\label{eer}
P_{fa}(\theta)&=\frac{\# \{{\textit{fake trials with score $\textgreater$ $\theta$}} \}}{\# \{{\textit{total fake trials}} \}} \\
P_{miss}(\theta)&=\frac{\# \{{\textit{genuine trials with score $\textless$ $\theta$}} \}}{\# \{{\textit{total genuine trials}} \}}
\end{align}
\vspace{-10pt}

So $P_{fa}(\theta)$ and $P_{miss}(\theta)$ are, respectively, monotonically decreasing and increasing functions of $\theta$. The EER corresponds to the error rate at the threshold $\theta_{EER}$ at which the two detection error rates are equal, i.e. $EER=P_{fa}(\theta_{EER}) =P_{miss}(\theta_{EER})$.

There are two rounds of evaluations in the detection task of audio fake game track in ADD challenges~\cite{Yi2022ADD, Yi2023ADD}. Each round evaluation has each own ranking in terms of EER. The final ranking is in terms of the weighted EER (WEER), which is defined as follows.
\vspace{-5pt}
\begin{eqnarray}\label{weer}
WEER= \alpha * EER\_{R1} + \beta * EER\_{R2}
\end{eqnarray}
where $\alpha$ and $\beta$ denotes the weight of the corresponding EER, $EER\_{R1}$ and $EER\_{R2}$ are the EER of the first and second round evaluation in the detection task of audio fake game track, respectively.

\begin{figure*}[htb]
\hfill
\begin{minipage}[b]{1.0\linewidth}
  \centering
  \centerline{\includegraphics[width=17.0cm,height=9.8cm]{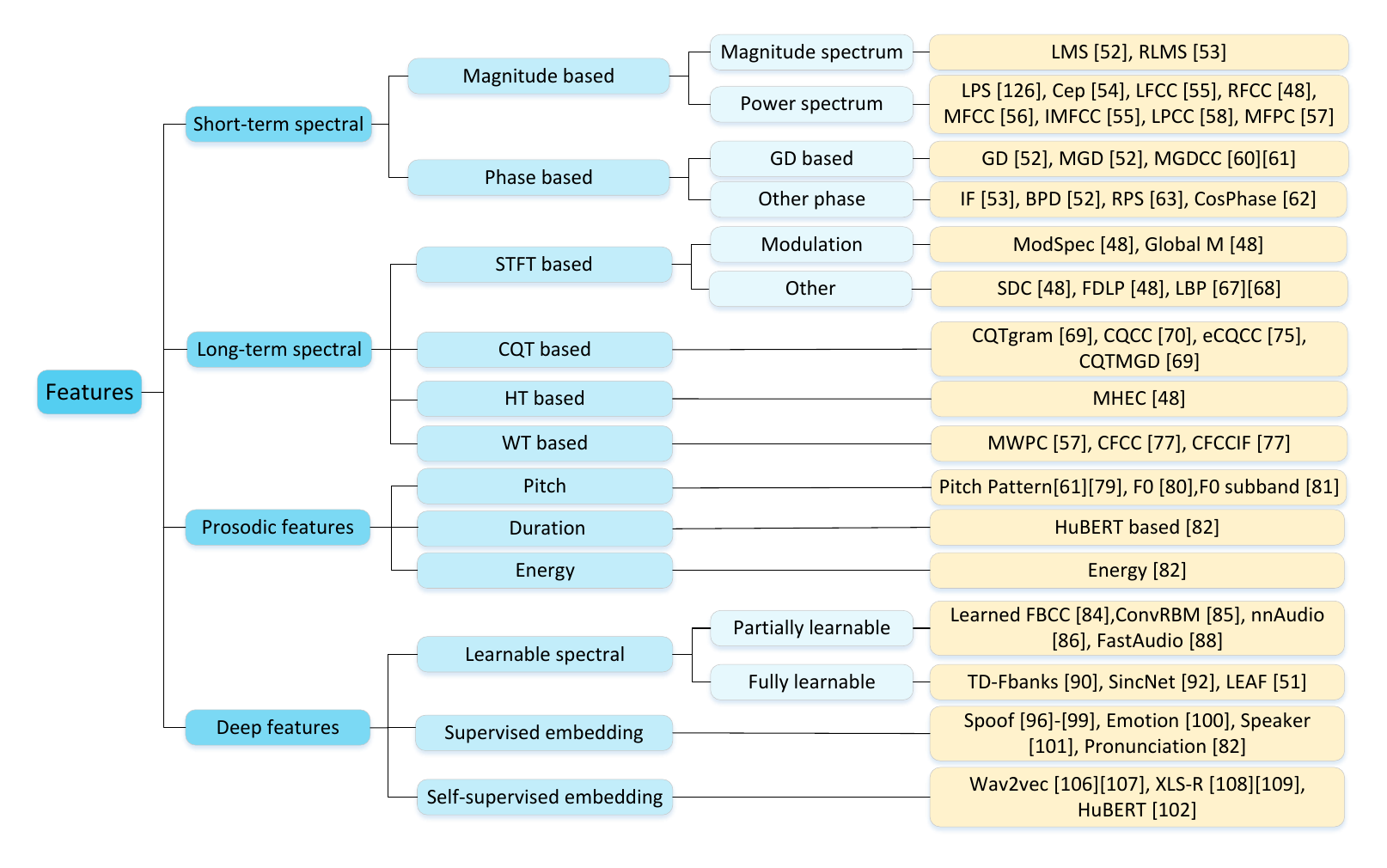}}
\end{minipage}
\caption{The typical features used in previous studies can be roughly divided into four categories: short-term spectral features, long-term spectral features, prosodic features and deep features.}
\label{fig:features}
\end{figure*}

\section{Discriminative Features}
The feature extraction is a key module of the pipeline detector. 
The goal of feature extraction is to learn discriminative features via capturing audio fake artifacts from speech signals. Large amounts of efforts~\cite{Sahi2015A,Das2019LongRA,Wang2021InvestSF} have shown the importance of useful features for detecting fake attacks. The features used in previous studies can be roughly divided into four categories: short-term spectral features, long-term spectral features, prosodic features and deep features.
Short- and long-term spectral features are extracted largely by relying on digital signal processing algorithms. Short-term spectral features, extracted from short frames typically with durations of 20-30 ms, describe the short-term spectral envelope involving an acoustic correlate of voice timbre. However, short-term spectral features have been demonstrated inadequate in capturing temporal characteristics of speech feature trajectories. In response to this, some researchers propose long-term spectral features to capture long-range information from speech signals. In addition, prosodic features are used to detect fake speech. Unlike the short-term spectral features from short duration, prosodic features spans over longer segments, such as phones, syllables, words, and utterances etc. Most of the aforementioned spectral and prosodic features are hand-crafted features, the design of which is flawed by biases due to limitations of handmade representations~\cite{2021LEAF}. So deep features, extracted via deep neural network based models, are motivated to fill the gap. The characteristics and relationships of different features are listed in Figure~\ref{fig:features}.

\subsection{Short-term Spectral Features}
Short-term spectral features are computed mainly by applying the short-time Fourier transform (STFT) on a speech signal~\cite{Xiao2015Spoofing}.
Given a speech signal $x(t)$, it is assumed to be quasi-stationary within a short period (e.g. 25ms).
The STFT of the speech signal $x(t)$ is formulated as follows:
\begin{equation}
X(t,\omega) = \left|X(t,\omega) \right|e^{j\phi(\omega)},
\end{equation}
where $\left|X(t,\omega) \right|$ is the magnitude spectrum and $\phi(\omega)$ is the phase spectrum at frame $t$ and frequency bin $\omega$. 
The power spectrum is defined to be $\left|X(t,\omega) \right|^{2}$. 

Short-term spectral features are mainly composed of short-term magnitude and phase based features.
Usually, few of the magnitude based features are directly derived from the magnitude spectrum but most of them are derived from the power spectrum. The phase based features are derived from the phase spectrum.

\subsubsection{Short-term magnitude based features}
The statistical averaging inherent in parametric modeling of the magnitude spectrum may introduce artefacts, such as over-smoothed spectral envelopes. The use of magnitude based spectrum can therefore be useful for detecting generated speech. Short-term magnitude features include magnitude spectrum and power spectrum features.

\textbf{Magnitude spectrum features} are directly derived from the magnitude spectrum~\cite{Xiao2015Spoofing, Tian2016Spoofing}. The logarithm of magnitude spectrum is called log magnitude spectrum (LMS) containing the formant information, harmonic structure and all the spectral details of speech signal. The logarithmic is used to reduce the the dynamic range of the magnitude spectrum. Formant information contained in LMS is important for speech recognition but may not be useful for fake detection as most of the fake techniques (e.g. TTS or VC) are effective in modelling the formant of speakers. Therefore, residual log LMS (RLMS) is proposed by employing inverse linear predictive coding (LPC) filter to reduce the impact of formant information but better analyse the details of spectrum such as harmonics.

\textbf{Power spectrum features} are derived from the power spectrum, which may be the most well studied in fake audio detection. They include log power spectrum (LPS), cepstrum (Cep), filter bank based cepstral coefficients (FBCC), all-pole modeling based cepstral coefficient (APCC) and subband spectral (SS) features. 
LPS, commonly called log-spectrum, is computed directly on raw power spectrum by the logarithm~\cite{Rabiner1999Fund}. 
Cep is derived from the power spectrum by applying discrete cosine transform (DCT). However, the dimensionality of LPS and Cep features is too high. 
FBCC features~\cite{Sahi2015A} are proposed to address the aforementioned issue, and include rectangular filter cepstral coefficients (RFCC), linear frequency cepstral coefficients (LFCC), mel frequency cepstral coefficient (MFCC), inverted MFCC (IMFCC). RFCC is computed using linear scale rectangular filters. LFCC~\cite{Todisco2018Int} is extracted with linear triangular filters. MFCC~\cite{Chen2010MFCC} is derived from mel scale triangular filters, with denser placement in lower-frequencies to simulate human-ear perception. IMFCC~\cite{Todisco2018Int} utilizes triangular filters that are linearly spaced on inverted-mel scale, giving higher emphasis to the high-frequency region. Mel-frequency principal coeffitients (MFPC) features~\cite{2016STC} are obtained similarly to the MFCC coefficients, but using principal component analysis (PCA) instead of the DCT to reomve the relations of the acoustic features. Mel spectrum (Mel spec) also is derived similarly to the MFCC coefficients without DCT. LFCC features are well-known, which together with Gaussian mixture model (GMM) and light convolutional neural network (LCNN) are used as the baseline models for ASVspoof~\cite{2021ASVspoof} and ADD~\cite{Yi2022ADD, Yi2023ADD} challenges.
APCC features~\cite{Sahi2015A} are derived from all-pole modeling representation of signal converted to linear prediction cepstral coefficients (LPCC)~\cite{Chak2008Improved}. 
SS features~\cite{Sahi2015A} include subband spectral flux coefficients (SSFC), spectral centroid magnitude coefficients (SCMC), subband centroid frequency coefficients (SCFC) and discrete Fourier mel subband transform (DF-MST)~\cite{2020Significance}. The subband features mostly extract information such as spectral flux and centroid magnitude without looking into the details within each subband.

\subsubsection{Short-term phase based features}
Even though phase information is important in human speech perception, most TTS and VC systems use a simplified, minimum phase model which may introduce artefacts into the phase spectrum. Therefore, phase based features can be used to discriminate between human and generated speech.
The phase spectrum itself does not have stable patterns for fake audio detection due to phase warping~\cite{Xiao2015Spoofing}. Post-processing methods are instead utilised to generate useful short-term phase based features including group delay (GD) based and other phase features.

\textbf{GD based features} involve GD, modified GD (MGD), MGD cepstral coefficients (MGDCC) and all-pole group delay (APGD).
GD is the derivative of phase spectrum along the frequency axis, which is referred as to a representation of filter phase response.
MGD is computed from the spectrum after cepstral smoothing frame-by-frame, which is a variation of GD and can extract a more clear phase pattern than GD. Xiao et al.~\cite{Xiao2015Spoofing} use two factors to control the dynamic range of the MGD for anti-spoofing.
MGDCC is computed from the MGD phase spectrum, using both phase and magnitude information~\cite{Wu2013Synthetic, Wu2016SAS}. Wu et al.~\cite{Wu2012Spoofing} use MGDCC features to distinguish between synthetic speech and human speech, which outperform MFCC features.
APGD is a phase-based feature using all-pole modeling, whose role in spoofed speech detection is investigated, notably in~\cite{Sahi2015A}, which has fewer parameters compared to MGD due to only the all-pole predictor order needing to be optimized.

\textbf{Other phase features}~\cite{Xiao2015Spoofing, Tian2016Spoofing} include instantaneous frequency (IF), baseband phase difference (BPD), relative phase shift (RPS), pitch synchronous phase (PSP) and cosine-phase (CosPhase) based features. IF features~\cite{Tian2016Spoofing} is the derivative of the phase spectrum along the time axis. Different from the raw phase spectrum that scarcely reflects any patterns, the IF spectrum capturing the temporal information of phase has clear patterns. The IF and GD contain very different patterns, which could provide complementary information for spoofed speech detection~\cite{Xiao2015Spoofing}.
BPD is a phase feature extracted from baseband STFT, which can provide more stable time-derivative phase information compared to the IF.
RPS~\cite{2015Toward} reflects the "phase shift" of harmonic components in relation to the fundamental frequency. 
Another way to reveal the patterns in phase spectrum is to use pitch synchronous STFT, where the patterns are called PSP features~\cite{Xiao2015Spoofing}.
CosPhase features~\cite{Wu2012Spoofing} are extracted from the phase spectrum by applying the cosine function to unwrapped phase spectrum followed by DCT. In order to reduce the dimensionality of CosPhase features, CosPhase principal coefficients (CosPhasePC)~\cite{2016STC} are computed by means of PCA.

\subsection{Long-term Spectral Features}
Short-term spectral features are not good at capturing temporal characteristics of speech feature trajectories due to being computed in a frame-by-frame fashion~\cite{Wu2013Synthetic}. Therefore, long-term spectral features have been proposed to capture long-range information from speech signals, and studies have shown that they are critical to fake speech detection~\cite{Das2019LongRA}.
The long-term features can be roughly categorized into four types in terms of time-frequency analysis approaches: STFT based features, constant-Q transform (CQT) based features, Hilbert transform (HT) based features and wavelet transform (WT) based features.

\subsubsection{STFT based features}

There are four kinds of STFT based features: modulation features, shifted delta coefficients (SDC), frequency domain linear prediction (FDLP) and local binary pattern (LBP) features.

Modulation features include modulation spectrum (ModSpec) and global modulation (Global M). ModSpec contains long-term temporal characteristics of speech signal~\cite{Sahi2015A}. Global M features combines spectral (e.g. MFCC) and temporal modulation information for better long range feature modeling to further improve the performance of fake audio detection~\cite{2021Generalized}. SDC captures long-term speech information and are computed by augmenting delta coefficients of multiple speech frames~\cite{Sahi2015A}. FDLP is obtained by performing DCT on speech signal using linear prediction analysis performed on different subbands, which are studied in fake audio detection~\cite{Sahi2015A}. LBP features obtain long span information upon spectral features via LBP analysis in computer vision tasks~\cite{Ojala2002Rota,Zhao2007Facial}. Alegre et al.~\cite{Alegre2013LBP, Alegre2013AOC} use uniform LBP analysis to convert LFCC features into a so-called textrogram, the histograms of which are used for spoof detection.

\subsubsection{CQT based features}
Unlike short-term spectral features derived from a window of tens of milliseconds, the CQT is a long-term window transform. 
The CQT provides higher frequency resolution at lower frequencies, but higher temporal resolution at higher frequencies in contrast to the STFT.
The center frequencies of each filter and the octaves are geometrically distributed for CQT. Various CQT based features are derived using CQT in different ways, which include CQT spectrum, CQ cepstral coefficient (CQCC), extended CQCC (eCQCC), inverted CQCC (ICQCC), and CQT-based modified group delay (CQTMGD) etc.

CQT spectrum is known as CQTgram~\cite{2019Replay}, which is computed by directly applied the logarithm on raw power magnitude spectrum obtained via CQT~\cite{Todisco2016Anew}. Lavrentyeva el al.~\cite{2017Audio} obtain the best results by using the CQT spectrum for audio replay attack detection of ASVspoof 2017.
CQCC is obtained from the DCT of the log power magnitude spectrum derived by CQT. In 2016, Todisco et al.~\cite{Todisco2016Anew} achieved promising performce of detecting unit selection TTS based attacks via CQCC features. CQCC features enjoy wide usage, including as input features of the baseline models of ASVspoof and ADD challenges~\cite{2021ASVspoof, Yi2022ADD}.  Over half of the ASVspoof 2019 participants (26 out of 48) ultilsed CQCC features as the input of their classifiers~\cite{Tak2020An}, many of which obtain top-performing results~\cite{2019Ensemble,2019STC}.
eCQCC is derived from the combination of coefficients from octave power spectrum with the CQCC features that are computed from linear power spectrum~\cite{2020Assessing}. 
ICQCC is derived from the inverted linear power spectrum of long-term CQT~\cite{2019Long}.
Cheng et al.~\cite{2019Replay} propose to incorporate the CQT and MGD for a more powerful representation of phase-based features named CQTMGD, which won the 1st place of the ASVspoof 2019 physical access sub-challenge.
 
\subsubsection{HT based features}
HT based features are computed from the analytical signal obtained by the HT, such as mean Hilbert envelope coefficients (MHEC)~\cite{Sahi2015A}. The Hilbert envelope is computed from each Gammatone filter output on the signal and a low-pass filter is used for smoothing.

\subsubsection{WT based features}
WT based features are derived mainly by performing WT on a speech sigal, which include mel wavelet packet coefficients (MWPC), cochlear filter cepstral coefficients (CFCC) and CFCC plus instantaneous frequency (CFCCIF).

MWPC is computed by performing wavelet-packet transform on speech signals. Novoselov et al.~\cite{2016STC} applied PCA to the log mel scale information to derive 12 coefficients, which are called MWPC features.
CFCC~\cite{2015Combining} is derived based on wavelet transform-like auditory transform, the relevant mechanism of which occurs in the cochlea of the human ear. 
CFCCIF denotes CFCC plus IF features at the output of each subband filters. The IF and phase of the envelope of the cochlear filter are vital features for speech perception of human listeners. TTS and VC models generate the speech in a frame-by-frame pattern. However, human speech production system does not produce speech at frame level rather in continuum. Therefore, Patel et al.~\cite{2015Combining} propose CFCCIF features with variation capturing across frames to discriminate the real speech from the spoofed one, which won the best result of ASVspoof 2015.

\subsection{Prosodic Features}
Prosody refers to non-segmental information of speech signals, including syllable stress, intonation patterns, speaking rate and rhythm~\cite{Tomi2010Overview}. Unlike the short-term spectral features from short duration typically of 20–30 ms, it spans over longer segments, such as phones, syllables, words, and utterances etc. 
The important prosodic parameters include fundamental frequency (F0), duration (e.g. phone duration, pause statistics), energy distribution, speaking rate etc. Previous studies~\cite{Tomi2010Overview} on fake audio detection mainly consider three major prosodic features: F0, duration and energy. These features are less sensitive to channel effects when compared to spectral features~\cite{Wu2015Spoofing}. They can provide complementary information to spectral features for improving the performance of fake audio detection.

F0 is also known as pitch. The pitch pattern of synthetic speech is different from that of natural speech~\cite{2012Synthetic}. It is difficult for TTS or VC models to precisely model human physiological features required to properly synthesize natural speech. So synthetic speech has a different mean pitch stability than human speech. In addition, co-articulation of human speech is  smoother and more relaxed than that of synthetic speech. This difference is captured by the jitters in the pitch pattern of the latter. Therefore, De Leon et al.~\cite{2012Synthetic} use pitch pattern statistics like mean pitch stability derived from image analysis for synthetic speech detection. Wu et al.~\cite{Wu2016SAS} introduce pitch pattern calculated by dividing the short-range autocorrelation function for anti-spoofing in 2016. Since TTS usually predicts F0 from text resulting in unnatural trajectories but VC usually copies a source speaker’s natural F0 trajectories, pitch pattern is more useful for detecting synthetic speech than VC speech, especially for unit selection synthesis attack. In 2018, Pal et al.~\cite{2018Synthetic} extracted pitch variation at frame-level as complementary information to magnitude and phase based features to improve the performance of synthetic speech detection. However, the distribution of F0 is irregular so that it is difficult to use it directly. In order to address this issue, Xue et al.~\cite{xue2022f0} propose a method to capture discriminative features of the F0 subband for fake speech detection, which not only uses the F0 information but also spectral features in 2022. However, pitch extraction algorithms are generally unreliable in noisy environments and the extraction of prosodic features requires relatively large amounts of training data due to their sparsity. Most recently, Wang et al.~\cite{Wang2023Prosody} first try to fuse F0, phoneme duration and energy for fake audio detection. Phoneme duration features are extracted from a pre-trained model HuBERT~\cite{Hsu2021HuBERT} trained using a large amount of speech data. 

\subsection{Deep Features}
The aforementioned spectral features and prosodic features are almost all hand-crafted features have strong and desirable representation abilities. However, their design is flawed by biases due to limitations of handmade representations~\cite{2021LEAF}. Therefore, deep features are motivated to fill in the gap. Deep features are learned by using deep neural networks, which can be roughly categorized into: learnable spectral features, supervised embedding features and self-supervised embedding features.

\subsubsection{Learnable spectral features}

 Learnable spectral features involve using learnable neural layers to estimate the standard filtering process, which can be categorized in terms of the procedures they perform: partially and fully learnable spectral features.

\textbf{Partially learnable spectral features} are extracted by training a neural network based filterbank matrix with a spectrogram obtained by applying STFT on a speech signal.
In 2017, Hong et al.~\cite{2017DNNFBank} developed deep neural network (DNN) filter bank cepstral coefficients (FBCC) named learned FBCC, to distinguish natural speech from spoofed one. The learned FBCC can capture better differences between real and synthetic speech than most hand-crafted FBCC, especially for detecting unseen attacks. Sailor et al.\cite{Sailor2017FbankCNN} propose a method to learn filterbank representation using convolutional restricted boltzmann machine named ConvRBM features.
In 2020, Cheuk et al.~\cite{2020nnAudio} presented a neural network-based audio processing toolkit named nnAudio, which uses 1D convolutional neural networks to transform audio signal from time domain to frequency domain~\cite{2013LearningFB}. This toolkit on make the waveform-to-spectram transformation layer trainable via back-propagation. However, Fu et al.~\cite{Fu2021FastAudioAL} report that nnAudio based anti-spoof methods obtain limited improvement due to the fact that nnAudio is implemented by a set of unconstrained learnable filterbanks. Zhang et al.\cite{Zhang2019DiscFF} use a neural network to learn the frequency centre, bandwidth, gain, and shape of the filter banks performing different constraints to extract features. Fu et al.~\cite{Fu2021FastAudioAL} propose a front-end named FastAudio whose input is a spectrogram of the STFT. The learnable layer of FastAudio is instead of replacing fixed filterbanks by performing filterbank shape constraints for anti-spoofing tasks. 

\textbf{Fully learnable spectral features} are learned directly from raw waveforms to approximate the standard filtering process. They are different from partially learnable spectral features extracted by training a filterbank matrix with a spectrogram. Zeghidour et al.~\cite{Zeg2018LearningFF} propose time-domain filterbanks (TD-FBanks) via scattering transform approximation of mel-filterbanks~\cite{Andn2013DeepSS}. The TD-FBanks are learned without any constraints to approximate mel-filterbanks at initialization. SincNet~\cite{Rav2018Sincnet} is proposed to learn a convolution with sine cardinal filters, a non-linearity and a max-pooling layer, as well as a variant using Gabor filters~\cite{No2020CGCNNCG}. Tak et al~\cite{Tak2021RawNet2} use SincNet as the first layer of the end-to-end anti-spoofing model called RawNet2. In 2021, Zeghidour et al.~\cite{2021LEAF} designed a new learnable filtering layer with Gabor filters called LEAF, which can be used as a drop-in substitute of mel-filterbanks. Unlike Sinc filters that require using a window function~\cite{Rav2018Sincnet}, Gabor filters are optimally localized in time and frequency domain. Tomilov et al.~\cite{Tomilov2021STCAS} obtain promising results by using LEAF features for detecting replay attacks of ASVspoof 2021~\cite{2021ASVspoof}.

\subsubsection{Supervised embedding features}

Supervised embedding features involve the extraction of deep embeddings from deep neural networks via supervised training. There are about four kinds of supervised embedding features for audio deepfake detection: spoof embeddings, emotion embeddings, speaker embedings and pronunciation embeddings.

\textbf{Spoof embeddings} are extracted from a neural network based model trained on the bonafide and spoofed data.  Chen et al.~\cite{Chen2015Svector} use a DNN based model to compute robust and abstract feature representation for spoofed speech detection in ASVspoof 2015 challenge. Qian et al.~\cite{Qian2016DeepFF} extract sequence-level bottleneck features, named s-vector, from recurrent neural network (RNN) models for anti-spoofing. Das et al.~\cite{Das2019LongRA} train a deep feature extractor using a DNN model with LPS features in 2019. The spoof embeddings are computed from the feature extractor by removing the output layer. In order to learn sequence contextual information for fake audio detection, Alejandro et al.~\cite{Alans2019ALC} propose a light convolutional gated RNN to learn utterance-level deep embeddings, which are then used as the inputs of the back-end classification. Most recently, Doan et al.~\cite{Doan2023BTSEAD} use RNN, convolutional sequence-to-sequence and transformer based encoder to learn breathing and talking sounds as well as silence in an audio clip for deepfake detection.

\textbf{Emotion embeddings} are learned using a supervised speech emotion recognition model trained with emotion labelled data. The emotion embeddings are directly used to detect fake utterances. Conti et al.~\cite{Conti22Emotion} proposed a method to detect fake speech via emotion recognition in 2022. The rationale behind this method is that the emotional behavior of the generated audio is not natural like that of real human speech. The results in~\cite{Conti22Emotion} show that the method can generalize well in cross-dataset scenarios.

\textbf{Speaker embeddings} are trained using a supervised speaker recognition model using training data with speaker identity label. The speaker embeddings are used as auxiliary features to improve the performance of fake audio detection. In 2022, Pan et al.~\cite{Pan2022Speaker} joinly train a speaker recognition model and a fake audio detection model via multi-objective learning. The speaker embeddings and LFCC features are both used as the input features of fake audio detection model.

\textbf{Pronunciation embeddings} are extracted from a speech recognition model trained with labelled data. The pronunciation embeddings can be directly used to discriminate the real speech from the fake one. In 2023, Wang et al.~\cite{Wang2023Prosody} fuse pronunciation embeddings and prosodic features to train a fake audio detection model. The pronunciation embeddings are computated from a pretrained conformer-based speech recognition model.

\subsubsection{Self-supervised embedding features}
Although supervised embedding features are well generalized to unknown conditions, they are learnt using a plentiful supply of labeled training data~\cite{Wang2021InSF}. However, obtaining annotated speech data or fake utterances is costly and technically demanding. This motivates researchers to extract deep embedding features from self-supervised speech model trained using any bona fide speech data. Despite training an effective self-supervised model costly, there are a number of pre-trained self-supervised speech models are publicly available, such as Wav2vec~\cite{Sch2019wav2vec, Baevski2020wav2vec2}, XLS-R~\cite{Babu2021XLSR} and HuBERT~\cite{Hsu2021HuBERT}. 

\textbf{Wav2vec based features} are extracted from the pretrained Wav2vec or Wav2vec2.0 models. Xie et al.~\cite{Xie2021Siamese} propose a Siamese neural network based representation learning model to distinguish real and fake speech in 2021. The model is trained with the wav2vec features extracted from a pretrained Wav2vec model. 
Tak et al.~\cite{Tak2022CMW2V2} use self-supervised learning in the form of a Wav2vec2.0 front-end with fine tuning for fake audio detection in 2022. Although the pretrained Wav2vec2.0 is trained using only genuine speech data without any fake audio, they obtain the state-of-the-art results reported in the literature for both the ASVspoof 2021 LA and Deepfake datasets.

\textbf{XLS-R based features} are extracted from the pre-trained XLS-R models which is a variant of Wav2vec2.0. Martin-Donas~\cite{Martin2022VA} utilize deep features extracted from pre-trained XLS-R~\cite{Babu2021XLSR}, which is a large-scale model for cross-lingual speech representation learning based on a pretrained Wav2vec2.0. The method ranked first in the LF track of ADD 2022 challenge~\cite{Yi2022ADD}, where the utterances are interfered with various noises. Lv et al.~\cite{Lv2022FakeAD} use a self-supervised model XLS-R as a feature extractor for fake audio detection.
The features generalize well for unknown partially fake voices and obtain the best results of PF task of ADD 2022 competition~\cite{Yi2022ADD}.

\textbf{HuBERT based features} are extracted from the pre-trained HuBERT models. Wang et.~\cite{Wang2023Prosody} use a pre-trained HuBERT~\cite{Hsu2021HuBERT} model to extract the duration encoding vector for audio deepfake detection. The encoding vector is an encoding similar to speech phonemes. Wang et al.~\cite{Wang2021InSF} directly use the embeddings from the pre-trained HuBERT as the input features of the detection models.

Wang et al.~\cite{Wang2021InSF} also investigate the performance of spoof speech detection using embedding features extracted from different pre-trained self-supervised models, e.g. Wav2vec2.0, XLS-R and HuBERT, providing some useful findings in~\cite{Wang2021InSF}. If the pre-trained model is not fine-tuned with target data, the classifier needs to be deep for the anti-spoofing models. However, a simple neural network with just an average temporal pooling and linear layer is sufficient when the pre-trained model is fine-tuned with anti-spoofing data. Learning deep embedding features using self-supervised training is suggested as a potential direction to improve the generalization of fake audio detection.

\begin{table*}[]
	\centering
	\caption{Comparison of representative audio deepfake detection classifications}
        \label{table:classification}
	\resizebox{\textwidth}{!}{
		\begin{tabular}{cclll}
			\toprule[1pt]
			\rowcolor[HTML]{FFFFFF} 
			\multicolumn{1}{l}{\cellcolor[HTML]{FFFFFF}} & \multicolumn{1}{l}{\cellcolor[HTML]{FFFFFF}} & Algorithm & Advantages & Disadvantages \\ \hline
			\rowcolor[HTML]{EFEFEF} 		
			\cellcolor[HTML]{EFEFEF} Traditional&  & SVM~\cite{Alegre2012Spoof, Alegre2013AOC, Lpez2015DNNSVM} & Early work, excellent classification capabilities & Restricted by the limited training samples \\
			\rowcolor[HTML]{FFFFFF} 
			\cellcolor[HTML]{EFEFEF} Classification & \cellcolor[HTML]{EFEFEF} & GMM~\cite{Amin2013Disguise, 2012Synthetic, Sizov2015SVIvec, Ji2017EnsembleLF, 2017Constant} & Most widely used baseline & Performance needs to be further improved \\ 
               \hline
			\rowcolor[HTML]{EFEFEF} 
			\cellcolor[HTML]{EFEFEF} & CNN based & LCNN~\cite{Hemavathi2021Voice, Todisco2019ASVspoof, Yi2022ADD, 2019STC, Zeinali2019VGG, Wu2020LCNN} & Easier to benchmark due to popularity & \begin{tabular}[c]{@{}l@{}}Deeper networks are difficult to train and result \\ in performance degradation\end{tabular}\\
			\rowcolor[HTML]{FFFFFF} 
			\cellcolor[HTML]{EFEFEF} & \cellcolor[HTML]{FFFFFF} & ResNet~\cite{Alzantot2019DeepRN, Tomilov2021STCAS, Chen2021Pind, Parasu2020Inves} & Avoiding performance degradation & Limitted generalizabilities to unseen fake attacks \\
			\rowcolor[HTML]{EFEFEF} 
			\cellcolor[HTML]{EFEFEF} & \multirow{-2}{*}{\cellcolor[HTML]{FFFFFF}ResNet based}& AFN~\cite{Lai2018AFN} & \begin{tabular}[c]{@{}l@{}}Enhances feature representations \\ in both the frequency and time domains \end{tabular} & Out-of-domain generalizabilities should be improved \\
			\rowcolor[HTML]{EFEFEF} 
			\cellcolor[HTML]{EFEFEF} & Res2Net based & Res2Net~\cite{Li2020Res2Net, Kim2022PhaseAware} & \begin{tabular}[c]{@{}l@{}}Enlarging the receptive fields and improve the \\ generalization to unseen fake utterance \end{tabular} & Not considering channel relationship \\
			\rowcolor[HTML]{FFFFFF} 
			\cellcolor[HTML]{EFEFEF} & \cellcolor[HTML]{FFFFFF} & SENet~\cite{Hu2017SENet, Zhang2021TheEO} & Modelling interdependencies between channels & \begin{tabular}[c]{@{}l@{}}Deeper networks are difficult to train and result in \\ performance degradation \end{tabular}\\
			\rowcolor[HTML]{EFEFEF} 
			\cellcolor[HTML]{EFEFEF} & \multirow{-2}{*}{\cellcolor[HTML]{FFFFFF}SENet based} & ASSERT~\cite{Lai2019ASSERTAW} & \begin{tabular}[c]{@{}l@{}}Combining SENet and ResNet, achieving better \\performance \end{tabular} & \begin{tabular}[c]{@{}l@{}}Not learning the relationships between \\ neighbouring sub-bands or segment \end{tabular} \\
			\rowcolor[HTML]{FFFFFF} 
			\cellcolor[HTML]{EFEFEF} & \cellcolor[HTML]{EFEFEF}GNN based & GAT~\cite{Tak2021GraphAN} & \begin{tabular}[c]{@{}l@{}}Modelling relationships between temporal \\ segments or spectral sub-bands \end{tabular}& Not automatically optimize network architectures \\
			\rowcolor[HTML]{EFEFEF} 
			\multirow{-15}{*}{\cellcolor[HTML]{EFEFEF}\begin{tabular}[c]{@{}l@{}}Deep Learning \\ Classification\end{tabular}} & \cellcolor[HTML]{FFFFFF}DARTS based & PC-DARTS~\cite{Ge2021PCDARTS} & \begin{tabular}[c]{@{}l@{}} Little human effort, automatically optimizes the \\ operations in network architecture blocks \end{tabular} & Difficult to train \\ \hline
			\rowcolor[HTML]{FFFFFF} 
			\cellcolor[HTML]{EFEFEF} & \cellcolor[HTML]{EFEFEF}CNN based & CRNNSpoof~\cite{Chintha2020CRNN} & \begin{tabular}[c]{@{}l@{}}Early end-to-end work \\ for audio deepfake detection \end{tabular}& Performance need be improved \\
			\rowcolor[HTML]{EFEFEF} 
			\cellcolor[HTML]{EFEFEF} & \cellcolor[HTML]{FFFFFF} & RawNet2~\cite{Tak2021RawNet2, Hansen2022AAML} & Widely used end-to-end model &\begin{tabular}[c]{@{}l@{}} Not optimize the parameters of the Sinc-conv during \\ training \end{tabular}\\
			\rowcolor[HTML]{FFFFFF} 
			\cellcolor[HTML]{EFEFEF} & \multirow{-2}{*}{\cellcolor[HTML]{FFFFFF}RawNet2 based} & TO-RawNet~\cite{Wang2023TORawnet} & \begin{tabular}[c]{@{}l@{}}Reducing the correlation between filters in the \\ Sinc-conv \end{tabular}& Not learning adjacent temporal relationships \\
			\rowcolor[HTML]{EFEFEF} 
			\cellcolor[HTML]{EFEFEF} & \cellcolor[HTML]{EFEFEF} & RawGAT-ST~\cite{Tak2021RawGAT} & \begin{tabular}[c]{@{}l@{}}Data boosting and augmentation technique \\ with spectro-temporal GAT \end{tabular}&\begin{tabular}[c]{@{}l@{}} Not considering two heterogeneous graphs via a \\ heterogeneous attention mechanism \end{tabular} \\
			\rowcolor[HTML]{FFFFFF} 
			\cellcolor[HTML]{EFEFEF} & \multirow{-2}{*}{\cellcolor[HTML]{EFEFEF}GNN based} & AASIST~\cite{Jung2022AASIST} & \begin{tabular}[c]{@{}l@{}}Modelling artefacts spanning temporal and spectral \\ segments with a heterogeneous attention mechanism \end{tabular}& Unreliably for unknown fake attacks \\
			\rowcolor[HTML]{EFEFEF} 
   			\cellcolor[HTML]{EFEFEF} & \multirow{-1}{*}{\cellcolor[HTML]{EFEFEF}DARTS based} & Raw PC-DARTS~\cite{Ge2021RawDARTS} & \begin{tabular}[c]{@{}l@{}}Little human effort \\directly upon the raw speech \end{tabular}& Not easy to train \\
			\rowcolor[HTML]{FFFFFF} 
			\cellcolor[HTML]{EFEFEF} & \cellcolor[HTML]{EFEFEF} & Rawformer~\cite{Liu2023Rawformer} & \begin{tabular}[c]{@{}l@{}} Use positional-related local and global dependency \\ for synthetic speech detection \end{tabular}& Acquire local dependency not well \\ 
			\rowcolor[HTML]{EFEFEF} 
			\multirow{-11}{*}{\cellcolor[HTML]{EFEFEF}\begin{tabular}[c]{@{}l@{}}End-to-End \\ Model\end{tabular}} & \multirow{-3}{*}{\cellcolor[HTML]{EFEFEF}Transformer based} & SE-Rawformer~\cite{Liu2023Rawformer} & \begin{tabular}[c]{@{}l@{}}Using squeeze-and-excitation operation  \\ to acquire local dependency \end{tabular}&  Computation costly\\
			\hline
		\end{tabular}
	}
\end{table*}
 
\section{Classification Algorithms}
The back-end classifier is also very important for audio deepfake detection, which aims to learn high-level feature representation of the front-end input features and model excellent discrimination capabilities. The classification algorithms are mainly divided into two categories: traditional and deep learning classification.

\subsection{Traditional Classification}
Many classic pattern classification approaches have been employed to detect fake speech, including logistic regression (LR)~\cite{Rod2020LR, Singh2021AIS}, probabilistic linear discriminant analysis (PLDA)~\cite{Prince2012Plda,Ji2017EnsembleLF}, random forest (RF)\cite{Ji2017EnsembleLF}, gradient boosting decision Tree (GBDT)\cite{Ji2017EnsembleLF}, extreme learning machine (ELM)~\cite{Rahmeni2020ELM}, k-nearest neighbor (KNN)~\cite{Singh2021AIS} and so on. The most widely used classifiers are the support vector machine (SVM)~\cite{Chang2011LIBSVM} and GMM~\cite{Reynolds1995SRGMM}.

\subsubsection{SVM based classifiers}
One of the extensively used traditional classifiers in previous early work for spoofing audio detetion is SVM due to its excellent classification capabilities. 
Alegre et al.~\cite{Alegre2012Spoof} suggest that SVM classifiers are inherently robust to artificial signal spoofing attacks. However, it is very difficult to know the exact nature of spoofing attacks in practical scenarios. Therefore, Alegre et al.~\cite{Alegre2013AOC} and Villalba et al.~\cite{Lpez2015DNNSVM} propose a one-class SVM classifier only trained using genuine utterances to classify real and fake voices, which generalizes well to unknown spoof attacks. 
Hanilçi et al.~\cite{Hanili2015Class} use i-vectors as the input features of SVM to discriminate the real utterance from the fake one. 

\subsubsection{GMM based classifiers}
Another conventional classifier well-known as GMM is widely used in fake audio detection as it is an effective generative model employed as the baseline model in a series of competitions, such as ASVspoof 2017~\cite{2017ASVspoof}, 2019~\cite{Todisco2019ASVspoof}, 2021~\cite{2021ASVspoof} and ADD 2022~\cite{Yi2022ADD}.
Amin et al.~\cite{Amin2013Disguise} train a GMM classifier fed with MFCC features to detect voice disguise from speech variability.
De Leon et al.~\cite{2012Synthetic} use a GMM classification to discriminate between human and synthetic voices.
Wu et al.~\cite{Wu2012Phase} propose a method which decided between real and converted speech by using log-scale likelihood ratio based upon the GMM model for real and converted speech.
Sizov et al.\cite{Sizov2015SVIvec} use i-vectors trained with GMM mean supervector to jointly perform VC attacks detection and speaker verification obtaining promising performance.
Many participants of ASVspoof 2015 and 2017 have obtained promising performance by adopting GMM for classifying genuine and spoofed speech~\cite{Ji2017EnsembleLF}. Sahidullah et al.~\cite{Sahi2015A} choose GMM classifiers for benchmarking of various features as it yields reasonably good accuracy in the ASVspoof 2015 dataset. In addition, Todisco et al.~\cite{2017Constant} use GMM in a standard 2-class classifier in which the classes correspond to genuine and spoofed speech.

\subsection{Deep Learning Classification}

The back-end classifications of the latest fake audio detection systems are mostly based on deep learning methods, which significantly outperform the SVM and GMM based classifiers due to their powerful modelling capabilities~\cite{Godoy2015DL}. The model architectures of back-end classification are generally based on convolutional neural network (CNN), deep residual network (ResNet), modified ResNet (Res2Net), squeeze-and-excitation networks (SENet), graph neural network (GNN), differentiable architecture search (DARTS) and Transformer.

\subsubsection{CNN based classifiers}
Since CNNs are good at capturing spatially-local correlation, CNN based classifiers have achieved promising performance~\cite{Hemavathi2021Voice}, such as light CNN (LCNN)~\cite{Wu2018lcnn} consisting of convolutional and max-pooling layers with Max-Feature-Map (MFM) activation. LCNN is used as the baseline model of the ASVspoof~\cite{Todisco2019ASVspoof} and ADD~\cite{Yi2022ADD} competitions. The best system in ASVspoof 2017~\cite{ 2019Replay} and the best single system in the LA task of ASVspoof 2019~\cite{2019STC} are also utilize LCNN for fake audio detection. The MFM activation of LCNN not only filters the noise effects (ambient noise, signal distortion, etc.), retains the core information, but also reduces the computational cost and storage space. Zeinali et al. ~\cite{Zeinali2019VGG} use VGG-like network comprising several convolutional and pooling layers followed by a statistics pooling and several dense layers to detect fake utterances. Wu et al.~\cite{Wu2020LCNN} propose a feature genuinization transformer with CNN trained only using genuine speech, and the outputs of this transformer are then fed into the LCNN based classifier. 

\subsubsection{ResNet based classifiers}
Although deep CNNs have achieved promising results for fake audio detection, deeper neural networks are more difficult to train and result in performance degradation. In order to address this problem, ResNet~\cite{He2015DeepRL} is introduced as the classifier~\cite{Alzantot2019DeepRN}, employing a residual mapping. Tomilov et al.~\cite{Tomilov2021STCAS} and Chen et al.~\cite{Chen2021Pind} both use ResNet as the classifier for audio deepfake detection and achieve promising results in the deepfake task of ASVspoof 2021. 
Yan et al.~\cite{Yan2022AudioDD} employ a standard 34-layer ResNet with multi-head attention pooling layer for detecting deepfake audio, ranking in the first place in the FG-D task of the ADD 2022. Lai et al.~\cite{Lai2018AFN} propose a ResNet-based classifier named Attentive Filtering Network (AFN) to further improve the performance. AFN is based on dilated residual network, using convolution layers instead of fully connected layers, and modifying the residual units by adding a dilation factor. A light ResNet based fake audio detection system is introduced by Parasu et al. ~\cite{Parasu2020Inves}, reducing network parameters to prevent overfitting. Kwak et al.~\cite{Kwak2021ResMax} introduce a compact network ResMax combining MFM activation and ResNet for improving the performance of spoofed audio detection system.

\subsubsection{Res2Net based classifiers}
Although ResNet based models have strong ability to capture fake cues, their generalizabilities to unseen fake attacks are still limited. Therefore, Li et al.~\cite{Li2020Res2Net} propose the incorporation of Res2Net in fake audio detection systems, where the feature maps within one ResNet block are splitted into to multiple channel groups linking by a residual-like connection. The connection of Res2Net enlarges the receptive fields and improve the generalization to unseen fake utterances. Kim et al.~\cite{Kim2022PhaseAware} utilize Res2Net and Phase network, fed with phase and magnitude features, for detecting fake audio.

\subsubsection{SENet based classifiers}
Convolution operators of CNN aim to fuse both spatial and channel-wise information within local receptive fields at each layer, not considering channel relationship. Hu et al.~\cite{Hu2017SENet} and Zhang et al.~\cite{Zhang2021TheEO} propose a focus Squeeze-and-Excitation network (SENet) adaptively modelling interdependencies between channels. Lai et al.~\cite{Lai2019ASSERTAW} use SENet to train one of the fake audio detection models. Furthermore, they propose a system named Anti-Spoofing with Squeeze-Excitation and Residual neTworks (ASSERT) combining SENet and ResNet. The ASSERT are ranked as one of the top performing systems in the two sub-challenges in ASVspooof 2019. Wu et al.~\cite{Wu2022Partially} use SENet with self-attention layer to detect partially fake audio, achieving top performance in ADD 2022 challenge. Xue et al.~\cite{Xue2023LearningFY} utilize SENet with efficient channel attention via self-distillation for fake speech detection.

\subsubsection{GNN based classifiers}
Graph neural networks (GNNs)~\cite{Scarselli2009TheGN}, like graph attention network (GAT) or graph convolutional network (GCN), are used to learn underlying relationships among data. The fake artefacts used to detect spoofing attacks are often located in specific temporal segments or spectral sub-bands. However, the aforementioned studies do not focus on learning the relationships between neighbouring sub-bands or segments. Tak et al.~\cite{Tak2021GraphAN} use a GAT to model these relationships to improve the  performance of fake audio detection systems. More recently, Chen et al.~\cite{Chen2023Grap} utilize GCN incorporating prior knowledge to learn spectro-temporal dependency information for anti-spoofing, which achieves promising performance on the ASVspoof 2019 LA dataset.

\subsubsection{DARTS based classifiers}

A particular variant of neural architecture search known as differentiable architecture search (DARTS)~\cite{Liu2018DARTSDA}, automatically optimizes the operations contained within architecture blocks, including convolutional, pooling, residual connections operations. Ge et al.~\cite{Ge2021PCDARTS} introduce a variant of DARTS known as partial channel connections (PC-DARTS) for audio deepfake detection. The PC-DARTS based model, with little human effort and containing 85\% fewer parameters than a Res2Net model, obtains competitive results compared to the best performing systems in previous studies. Wang et al.~\cite{Wang2022LightDARTS} propose a light DARTS, which combines DARTS with MFM activation, playing the role of feature selection.

\subsubsection{Transformer based classifiers}
Different from the fully fake utterances, the partially fake utterances contains some discontinuity artifacts between concatenated audio clips. Transformer is good at modelling local and global artifacts and relationship. So Cai et al.~\cite{Cai2022WaveformBD} use Transformer and ResNet-1D as the backend classifier to detect partially fake audio and locate the fake regions.

\section{End-to-End Models}
The aforementioned methods to audio deepfake detection have focused on the design of machine learning based classifiers fed with hand-crafted or learnable features. 
Although past literature~\cite{2019STC, Li2020Res2Net, Tak2021GraphAN, Wang2021ACS} shows that the use of well-designed classifier usually leads to better performing models, the performance of a given classifier can vary greatly when combined with different features. In recent years, deep neural network based approaches that integrate feature extraction and classification in an end-to-end manner have achieved competitive performance, where both the feature extractor and the classifier are jointly optimized directly upon the raw speech waveform. The end-to-end models avoid limitations introduced from the use of knowledge-based features and are optimized for the application rather than generic decompositions~\cite{Tak2021RawNet2}. The end-to-end architectures of audio deepfake detection can be roughly classified into four types: CNN, RawNet2, ResNet, GNN, DARTS and Transformer.

\subsection{CNN based models}
Some researchers attempt the CNN based models to end-to-end fake audio detection. Muckenhirn et al.~\cite{Muck2017EtECNN} employed a simple CNN-based end-to-end approach to detection spoofed attacks. The proposed model consists of a single convolution layer and a multilayer perceptron (MLP) layer,  which performs well for VC and TTS attacks. A raw waveform convolutional long short term neural network (CLDNN) based anti-spoofing method is proposed by Dinkel et al.~\cite{Dinkel2017CLDNN}. The CLDNN model employs time- and frequency-convolutional layers to reduce time and spectral variations, as well as long-term temporal memory layers to model long-term temporal information.
In 2020, Chintha et al.~\cite{Chintha2020CRNN} proposed a convolution-recurrent neural network for spoofing detection named CRNNSpoof, which is composed of five 1-D convolution layers, a bidirectional LSTM layer and two fully-connected layers. However, the aforementioned models do not perform well in cross-dataset evaluation. In order to alleviate this issue, Hua et al. ~\cite{Hua2021TSSDNet} propose a time-domain synthetic speech section net, called TSSDNet, including Inception parallel convolutions structures named Inc-TSSDNet. The proposed model has promising generalization capability to unseen datasets. 

\subsection{RawNet2 based models}
Motivated by the power of RawNet2 in text-independent speaker verification~\cite{Jung2020Rawnet2}, Tak et al.~\cite{Tak2021RawNet2} employ RawNet2 to anti-spoofing. RawNet2 is a convolutional neural network with residual blocks, the first layer of which has a bank of sinc-shaped filters, which is essentially the same as that of SincNet ~\cite{Rav2018Sincnet}. RawNet2 operates directly on raw audio through time-domain convolution and has potential to learn cues that are not detectable using knowledge-based methods. Wang et al.~\cite{Hansen2022AAML} use RawNet2 with weighted additive angular margin loss for fake audio detection. However, RawNet2 does not optimize the parameters of the Sinc-Conv layer during training, limiting its performance. In order to alleviate this problem, Wang et al.~\cite{Wang2023TORawnet} propose TO-RawNet to improve its discriminability, which incorporates orthogonal convolution into RawNet2 reducing the correlation between filters in the sinc-conv. TO-RawNet based fake audio detection models observably outperform RawNet2 based models.

\subsection{ResNet based models}
Deeper neural network with residual mapping named ResNet become easy to train and achieve promising performance. Hua et al. ~\cite{Hua2021TSSDNet} propose a TSSDNet with residual skip connection named Res-TSSDNet, obtaining better performance. Ma et al.~\cite{Ma2021RWResnet} propose a speech anti-spoofing model named RW-ResNet composed of Conv1D Resblocks and backbone ResNet34. 

\begin{table*}[]
  \caption{Features and classifiers of the top-3 submitted systems for each task in ASVspoof and ADD competitions. The performance of each independent task is evaluated in terms of the EER (\%).}
  \label{table:topsystems}
	\resizebox{\textwidth}{!}{
	\begin{tabular}{cccccccccccc}
		\toprule[1pt]
		&  &  & \multicolumn{3}{c}{\textbf{Top 1}} & \multicolumn{3}{c}{\textbf{Top 2}} & \multicolumn{3}{c}{\textbf{Top 3}} \\
		\cmidrule(r){4-6} \cmidrule(r){7-9} \cmidrule(r){10-12}
		\multirow{-2}{*}{\textbf{Competition}} & \multirow{-2}{*}{\textbf{Year}} & \multirow{-2}{*}{\textbf{Task}} & \multicolumn{1}{l}{\textbf{EER (\%)}} & \textbf{Features} & \textbf{Classifiers} & \multicolumn{1}{l}{\textbf{EER (\%)}} & \textbf{Features} & \textbf{Classifiers} & \multicolumn{1}{l}{\textbf{EER (\%)}} & \textbf{Features} & \textbf{Classifiers} \\ \hline
		& \cellcolor[HTML]{EFEFEF}  & \cellcolor[HTML]{EFEFEF} & \cellcolor[HTML]{EFEFEF} & \cellcolor[HTML]{EFEFEF}MFCC, & \cellcolor[HTML]{EFEFEF} & \cellcolor[HTML]{EFEFEF}& \cellcolor[HTML]{EFEFEF}MFCC, MFPC, & \cellcolor[HTML]{EFEFEF}& \cellcolor[HTML]{EFEFEF}& \cellcolor[HTML]{EFEFEF} & \cellcolor[HTML]{EFEFEF}Mahalanobis \\
		& \multirow{-2}{*}{\cellcolor[HTML]{EFEFEF}2015~\cite{Sahidullah2019Intro}} & \multirow{-2}{*}{\cellcolor[HTML]{EFEFEF}LA} & \multirow{-2}{*}{\cellcolor[HTML]{EFEFEF}1.21} & \cellcolor[HTML]{EFEFEF}CFCCIF & \multirow{-2}{*}{\cellcolor[HTML]{EFEFEF}GMM} & \multirow{-2}{*}{\cellcolor[HTML]{EFEFEF}1.97} & \cellcolor[HTML]{EFEFEF}CosPhase & \multirow{-2}{*}{\cellcolor[HTML]{EFEFEF}SVM} & \multirow{-2}{*}{\cellcolor[HTML]{EFEFEF}2.53} & \multirow{-2}{*}{\cellcolor[HTML]{EFEFEF} s-vector} & \cellcolor[HTML]{EFEFEF}Distance \\
		& & & & LPS, & LCNN, & & CQCC, MFCC, &  GMM, GBDT, & & MFCC, IMFCC, PLP, & GMM \\
		& \multirow{-2}{*}{2017~\cite{Sahidullah2019Intro}} & \multirow{-2}{*}{PA} & \multirow{-2}{*}{6.73} & LPCC & GMM & \multirow{-2}{*}{12.34} & PLP & SVM, RF & \multirow{-2}{*}{14.03} & LFCC, RFCC, CQCC &  ANN \\
		& \cellcolor[HTML]{EFEFEF} & \cellcolor[HTML]{EFEFEF} & \cellcolor[HTML]{EFEFEF} & \cellcolor[HTML]{EFEFEF} & \cellcolor[HTML]{EFEFEF}ResNet,& \cellcolor[HTML]{EFEFEF} & \cellcolor[HTML]{EFEFEF}LFCC, CQT, & \cellcolor[HTML]{EFEFEF} & \cellcolor[HTML]{EFEFEF}& \cellcolor[HTML]{EFEFEF}MFCC, IMFCC, &\cellcolor[HTML]{EFEFEF}GMM, SVM \\
		& \multirow{-2}{*}{\cellcolor[HTML]{EFEFEF}} & \multirow{-2}{*}{\cellcolor[HTML]{EFEFEF}LA} & \multirow{-2}{*}{\cellcolor[HTML]{EFEFEF}0.22} & \multirow{-2}{*}{\cellcolor[HTML]{EFEFEF} Cep} & \multicolumn{1}{c}{\cellcolor[HTML]{EFEFEF}MobileNet}  & \multirow{-2}{*}{\cellcolor[HTML]{EFEFEF}1.86} & \cellcolor[HTML]{EFEFEF}LPS & \multirow{-2}{*}{\cellcolor[HTML]{EFEFEF}LCNN} & \multirow{-2}{*}{\cellcolor[HTML]{EFEFEF}2.64} & \cellcolor[HTML]{EFEFEF}SCMC, CQCC, Raw & \cellcolor[HTML]{EFEFEF} CNN, CRNN \\
		& \cellcolor[HTML]{EFEFEF} &  &  & CQT, MGD, &  &  & GD, LFCC, &  & & CQT, LFCC, & \\
		& \multirow{-4}{*}{\cellcolor[HTML]{EFEFEF}2019~\cite{Nautsch2021ASVspoof2S}} & \multirow{-2}{*}{PA} & \multirow{-2}{*}{0.39} &  CQTMGD & \multirow{-2}{*}{ResNet} & \multirow{-2}{*}{0.54} & Cep, IMFCC & \multirow{-2}{*}{ResNet} & \multirow{-2}{*}{0.59} & Cep & \multirow{-2}{*}{LCNN} \\
		& &\cellcolor[HTML]{EFEFEF} &\cellcolor[HTML]{EFEFEF} &  \cellcolor[HTML]{EFEFEF}Mel spec, & \cellcolor[HTML]{EFEFEF}LCNN, & \cellcolor[HTML]{EFEFEF}& \cellcolor[HTML]{EFEFEF} & \cellcolor[HTML]{EFEFEF} &\cellcolor[HTML]{EFEFEF} & \cellcolor[HTML]{EFEFEF} &\cellcolor[HTML]{EFEFEF} ResNet, \\
		&  & \multirow{-2}{*}{\cellcolor[HTML]{EFEFEF}LA} & \multirow{-2}{*}{\cellcolor[HTML]{EFEFEF}1.32} & \cellcolor[HTML]{EFEFEF}SincNet,Raw &  \cellcolor[HTML]{EFEFEF}ResNet & \multirow{-2}{*}{\cellcolor[HTML]{EFEFEF}2.77} & \multirow{-2}{*}{\cellcolor[HTML]{EFEFEF}LFCC} & \multirow{-2}{*}{\cellcolor[HTML]{EFEFEF}ResNet} & \multirow{-2}{*}{\cellcolor[HTML]{EFEFEF}3.13} & \multirow{-2}{*}{\cellcolor[HTML]{EFEFEF}Mel Spec} & \cellcolor[HTML]{EFEFEF}SENet \\
		& & & &  & VAE, & & LEAF,  & LCNN, & &  &  \\
		&  & \multirow{-2}{*}{PA} & \multirow{-2}{*}{24.25} & \multirow{-2}{*}{LPS} &  GMM & \multirow{-2}{*}{26.42} &  Mel spec & ResNet & \multirow{-2}{*}{27.59} & \multirow{-2}{*}{SincNet} & \multirow{-2}{*}{ResNet} \\
		& &\cellcolor[HTML]{EFEFEF} &\cellcolor[HTML]{EFEFEF} & \cellcolor[HTML]{EFEFEF}Mel spec,& \cellcolor[HTML]{EFEFEF}LCNN, &\cellcolor[HTML]{EFEFEF} & \cellcolor[HTML]{EFEFEF} & \cellcolor[HTML]{EFEFEF}ResNet, &\cellcolor[HTML]{EFEFEF} & \cellcolor[HTML]{EFEFEF} & \cellcolor[HTML]{EFEFEF} \\
		\multirow{-14}{*}{ASVspoof} & \multirow{-6}{*}{2021~\cite{Liu2022ASVspoof2T}} & \multirow{-2}{*}{\cellcolor[HTML]{EFEFEF}DF} & \multirow{-2}{*}{\cellcolor[HTML]{EFEFEF}15.64} & \cellcolor[HTML]{EFEFEF}SincNet, Raw & \cellcolor[HTML]{EFEFEF}ResNet & \multirow{-2}{*}{\cellcolor[HTML]{EFEFEF}16.05} & \multirow{-2}{*}{\cellcolor[HTML]{EFEFEF}Fbank} &  \cellcolor[HTML]{EFEFEF}MLP & \multirow{-2}{*}{\cellcolor[HTML]{EFEFEF}18.30} & \multirow{-2}{*}{\cellcolor[HTML]{EFEFEF}CQT} & \multirow{-2}{*}{\cellcolor[HTML]{EFEFEF}LCNN} \\ \hline
		& \cellcolor[HTML]{EFEFEF} &  &  & &  &  &  & ResNet, && CQT, &LCNN, \\
		& \cellcolor[HTML]{EFEFEF} & \multirow{-2}{*}{LF} & \multirow{-2}{*}{21.70} & \multirow{-2}{*}{XLS-R} & \multirow{-2}{*}{DNN} & \multirow{-2}{*}{23.00} & \multirow{-2}{*}{Log Fbank} & SENet & \multirow{-2}{*}{23.80} & Mel spec & ResMax \\
		& \cellcolor[HTML]{EFEFEF} & \cellcolor[HTML]{EFEFEF} & \cellcolor[HTML]{EFEFEF} & \cellcolor[HTML]{EFEFEF} & \cellcolor[HTML]{EFEFEF} & \cellcolor[HTML]{EFEFEF} & \cellcolor[HTML]{EFEFEF} & \cellcolor[HTML]{EFEFEF}SENet, & \cellcolor[HTML]{EFEFEF}& \cellcolor[HTML]{EFEFEF}Wav2Vec, &\cellcolor[HTML]{EFEFEF} ResNet, BLSTM, \\
		& \cellcolor[HTML]{EFEFEF} & \multirow{-2}{*}{\cellcolor[HTML]{EFEFEF}PF} & \multirow{-2}{*}{\cellcolor[HTML]{EFEFEF}4.80} & \multirow{-2}{*}{\cellcolor[HTML]{EFEFEF}XLS-R} & \multirow{-2}{*}{\cellcolor[HTML]{EFEFEF}TDNN} & \multirow{-2}{*}{\cellcolor[HTML]{EFEFEF}7.90} & \multirow{-2}{*}{\cellcolor[HTML]{EFEFEF}Mel spec} & \cellcolor[HTML]{EFEFEF} Self-attention & \multirow{-2}{*}{\cellcolor[HTML]{EFEFEF}9.40} & \cellcolor[HTML]{EFEFEF} Fbank & \cellcolor[HTML]{EFEFEF}Transformer \\
		& \cellcolor[HTML]{EFEFEF} &  &  & LFCC, &  &  & LFCC, CQT, & ResNet, & & LPS, & \\
		& \multirow{-6}{*}{\cellcolor[HTML]{EFEFEF}2022~\cite{Yi2022ADD}} & \multirow{-2}{*}{FG-D} & \multirow{-2}{*}{10.10} & Fbank & \multirow{-2}{*}{ResNet} & \multirow{-2}{*}{10.40} & Wav2vec2.0 & GMM, MLP & \multirow{-2}{*}{10.60} & CQT & \multirow{-2}{*}{LCNN}\\	
		& &\cellcolor[HTML]{EFEFEF} &\cellcolor[HTML]{EFEFEF} &\cellcolor[HTML]{EFEFEF} & \cellcolor[HTML]{EFEFEF} & \cellcolor[HTML]{EFEFEF}& \cellcolor[HTML]{EFEFEF} & \cellcolor[HTML]{EFEFEF}AASIST, &\cellcolor[HTML]{EFEFEF} & \cellcolor[HTML]{EFEFEF}CQT, &\cellcolor[HTML]{EFEFEF}LCNN, AASIST,  \\
		\multirow{-8}{*}{ADD} & \multirow{-2}{*}{2023~\cite{Yi2023ADD}} &  \multirow{-2}{*}{\cellcolor[HTML]{EFEFEF}FG-D} & \multirow{-2}{*}{\cellcolor[HTML]{EFEFEF}12.45} & \multirow{-2}{*}{\cellcolor[HTML]{EFEFEF}Wav2vec2.0} & \multirow{-2}{*}{\cellcolor[HTML]{EFEFEF}AASIST} & \multirow{-2}{*}{\cellcolor[HTML]{EFEFEF}17.93} & \multirow{-2}{*}{\cellcolor[HTML]{EFEFEF}LPS} & \cellcolor[HTML]{EFEFEF}SENet, LCNN & \multirow{-2}{*}{\cellcolor[HTML]{EFEFEF}22.13} & \cellcolor[HTML]{EFEFEF}Wav2vec2.0 &  \cellcolor[HTML]{EFEFEF}GMM \\ \bottomrule[1pt]
	\end{tabular}
}
\end{table*}

\subsection{GNN based models}
Inspired by the success of GAT to model complicated relationships among graph representations~\cite{Tak2021GraphAN}, Tak et al.~\cite{Tak2021RawGAT} propose a spectro-temporal GAT named RawGAT-ST, which learns the relationship, outperforming the RawNet2 and Res-TSSDNet on the LA evaluation set of ASVspoof 2019. However, the RawGAT-ST consists of a pair of parallel graphs to combine information by employing element-wise multiplication to the two graphs. In fact, it will be beneficial to combine the two heterogeneous graphs via a heterogeneous attention mechanism. Therefore, Jung et al.~\cite{Jung2022AASIST} propose a heterogeneous stacking graph attention layer to model artefacts spanning temporal and spectral segments with a heterogeneous attention mechanism, named AASIST. The AASIST outperforms the current state-of-the-art end-to-end models. Furthermore, a proposed lightweight variant called AASIST-L obtains competing performances~\cite{Jung2022AASIST}. These methods perform reliably under seen encoding and transimission conditions but unreliably for unknown telephony scenarios. In order to alleviate the problem, Tak et al.\cite{Tak2022Rawboost} propose a model based on RawGAT-ST and RawNet2 systems, named RawBoost, which is a data boosting and augmentation technique based upon the combination of linear and non-linear convolutive noises as well as impulsive and stationary additive noises that can be applied directly to raw audio. In addition, the AASIST does not optimize the parameters of the sinc-conv during training, which limited its performance. Therefore, Wang et al.~\cite{Wang2023TORawnet} employ orthogonal regularization in the Sinc-conv layer of the AASIST, which is called Orth-AASIST, outperforming the AASIST based model.

\subsection{DARTS based models}
The aforementioned end-to-end methods are encouraging and promising.  However, they can only automatically learn features and network parameters rather than network architecture. Therefore, Ge et al.~\cite{Ge2021RawDARTS} try to employ an automatic approach, which not only operates directly upon the raw speech signal but also  jointly optimizes of both the network architecture and network parameters. The appoarch is implemented based upon partially-connected differentiable architecture search from the raw audio waveform (Raw PC-DARTS). 

\subsection{Transformer based models}
In order to modelling local and global artefacts and relationship directly on raw audio, Liu et al.~\cite{Liu2023Rawformer} propose a model named Rawformer composed of convolution layer and transformer to detect fake utterances. The Rawformer generalizes better than the AASIST on cross-dataset evaluation. Liu et al.~\cite{Liu2023Rawformer} also propose a squeeze-and-excitation Rawformer called SE-Rawformer using squeeze-and-excitation operation to acquire local dependency, which outperforms the Rawformer.  

\section{Generalization Methods}

Although most of the existing audio deepfake detection methods have achieved impressive performance in in-domain test, their performance drops sharply when dealing with out-of-domain dataset in real-life scenarios~\cite{Korshunov2016Cross, Mller2022DoesAD, Chen2020General}. In other words, the generalization ability of audio deepfake detection systems is still poor. Several attempts have been made to try to tackle this challenge from different perspectives, such as loss function and continual learning.

\subsection{Loss Function}

It has become increasingly challenging to improve the generalization ability of audio deepfake detection systems to unknown attacks. In order to overcome this problem, Chen et al.~\cite{Chen2020General} ensure the neural network to learn more robust feature embeddings using large margin cosine loss (LMCL) function and online frequency masking augmentation. The generalization ability of detection models is increased by using LMCL and applying data augmentation. 
Zhang et al.~\cite{Zhang2020OneClass} use one-class learning to deal with unknown fake attacks, the key idea of which is to construct a compact representation of genuine audio representation and utilize an angular margin to separate the fake utterances in the embedding space. This method outperforms all previous existing single systems on the evaluation set of the LA task in ASVspoof 2019 challenge, without any data augmentation methods. These methods address the difficulties, to some degree, in detecting unknown attacks in practical use. However, the compactness of bona fide utterances in the embedding space lacks consideration of the diversity of speakers. Ding et al.~\cite{Ding2023SAMOSA} propose speaker attractor multi-center one-class learning (SAMO) to address the problem. The core idea of the SAMO is that real utterances are clustered around a number of speaker attractors and the method pushes away fake voices from all the attractors in a high-dimensional embedding space.

\subsection{Continual Learning}

Continual learning focuses on the continuous training and adaptation of models on new information, aiming to overcome catastrophic forgetting existing in fine-tuning. In order to improve the performance to unseen deepfake audio, Ma et al.~\cite{Ma2021Continual} propose a regularization based continual learning method, named Detecting Fake Without Forgetting (DFWF), to make the model learn new fake attacks incrementally. This method doesn't need to access old data but can ensure the model remember previous information. It also improves the detection performance on the new dataset and overcomes catastrophic forgetting by introducing regularization. However, the approximation of the DFWF may result in error accumulation in continual learning, leading to deteriorating learning performance. Most recently, Zhang et al.~\cite{Zhang2023RAWM} propose a continual learning algorithm for fake audio detection to solve this problem, called regularized adaptive weight modification (RAWM). RAWM relaxes the regularized constraint in the DFWF and introduce a adaptive direction modification to overcome catastrophic forgetting, and outperforms most of the typical continual learning methods in audio deepfake detection task.

\section{Performance Comparisons}

\subsection{Top-performing methods in typical competitions}

The top-performing systems in the ASVspoof and ADD competitions are summarised in Table~\ref{table:topsystems}, with their performance evaluated in terms of the EER. All of these systems use some form of data augmentation in their training, though no common form can be identified. Most are ensemble systems, and most operate upon short-term spectral features or raw waveforms. Most use a ResNet classifier or a variant thereof, with other types of convolutional networks also being popular. Fusion strategies employed in these methods include weighted averaging, using either uniformly or empirically set weights.

The systems generally rely on spectral features, especially in older iterations of the ASVspoof challenge. Recently, features derived from pre-trained models such as Wav2vec2.0 or XLS-R, have also seen much usage, with considerable success, as evidenced by their use in the top-performing systems in the ADD competitions.

In terms of classifiers, both traditional classifiers (e.g., SVMs and GMMs) and deep learning networks have been used in the top-performing systems. The top-performing systems in the ASVspoof competitions have used GMMs, SVMs, and ResNet classifiers, while the top-performing systems in the ADD competitions have used ResNet, LCNN and AASIST classifiers.

While superficially the performances seem to be decreasing, as seen from the increase in EER, this is likely due to the increasing difficulty of the tasks rather than a decrease in performance. In the competitions, the top-performing systems have consistently outperformed the baseline systems by a solid margin.
  
\subsection{Evaluation of Features}

We evaluate the discriminative performance of different handcrafted features in fake audio detection. The features include short-time spectral features (MFCC, LPS, LFCC, IMFCC), long-time spectral feature (CQCC), prosodic features (F0, energy, duration) and self-supervised embedded feature (XLS-R). We also include features resulting from concatenation of prosody features and LFCC or XLS-R. In this group of the experiments, we choose GMM, LCNN and ASSERT classifiers due to their popularity and effectiveness in fake audio detection tasks.

For the extraction of the short-time spectral features, we refer to the setups in the baseline systems. We then take the first- and second-order differences. For LFCC\footnote{\url{https://github.com/asvspoof-challenge/2021/tree/main/LA/Baseline-LFCC-LCNN}} and CQCC\footnote{\url{https://github.com/asvspoof-challenge/2021/tree/main/LA/Baseline-CQCC-GMM}}, we follow the default setup in the baseline systems. The LPS~\cite{Zhang2021TheEO} features are extracted with Blackman windows in STFT using the audio processing toolkit\footnote{\url{https://librosa.org/doc/latest/generated/librosa.stft.html}}.

For prosodic features, we use the Yet Another Algorithm for Pitch Tracking method (YAPPT)~\cite{zahorian2002yet} to extract the F0 features (A). The window length is set to be 25~ms with a window shift of 10~ms to extract the energy features (B). For duration features (C), we follow the setup of the HuBERT based duration method~\cite{Wang2023Prosody}.

We employ a variant of Wav2vec2.0, known as ``Wav2vec2.0 XLS-R\footnote{\url{https://github.com/facebookresearch/fairseq/tree/main/examples/wav2vec/xlsr}}'', to extract the XLS-R features. The model is pretrained on a dataset with 53 languages and 56 thousand hours of audio, and incorporates more linear transformations and a larger context network. A 10~ms audio segment is transformed to a 1024-dimensional vector for the pretrained model. 

\begin{table}[]
	\centering
	  \caption{The performance of representative features is evaluated using the detection models trained with the training set of ASVspoof 2021 in terms of the EER (\%). All the models are evaluated on the test set of DF task in ASVspoof 2021 and the test set In-the-Wild, respectively. }
        \label{table:features-asv}
        \fontsize{6pt}{8}\selectfont
	\resizebox{0.5\textwidth}{!}{
	\begin{tabular}{ccccccc}
		\toprule[1pt]
		\rowcolor[HTML]{FFFFFF} 
		\textbf{Features} & \multicolumn{3}{c}{\cellcolor[HTML]{FFFFFF}\textbf{ASVspoof 2021 DF}} & \multicolumn{3}{c}{\cellcolor[HTML]{FFFFFF}\textbf{In-the-Wild}} \\ 
		\cmidrule(r){2-4} \cmidrule(r){5-7}
		\rowcolor[HTML]{FFFFFF} 
		\textbf{} & \textbf{GMM} & \textbf{LCNN} & \textbf{ASSERT} & \textbf{GMM} & \textbf{LCNN} & \textbf{ASSERT} \\ \hline
		\rowcolor[HTML]{EFEFEF} 
		MFCC & 28.32 & 26.70 & 29.98 & 53.81 & 49.63 & 42.97 \\
		\rowcolor[HTML]{FFFFFF} 
		LPS & 28.64 & 31.96 & 24.67 & 65.38 & 83.63 & 35.32 \\
		\rowcolor[HTML]{EFEFEF} 
		LFCC & 25.25 & 33.74 & 39.67 & 37.49 & 35.14 & 48.82 \\
		\rowcolor[HTML]{FFFFFF} 
		IMFCC & 33.18 & 46.62 & 29.19 & 64.22 & 51.23 & 75.64 \\
		\rowcolor[HTML]{EFEFEF} 
		CQCC & 25.56 & 35.49 & 30.23 & 55.24 & 79.87 & 37.58 \\
		\rowcolor[HTML]{FFFFFF} 
		F0 (A) & 41.66 & 38.29 & 47.43 & 66.52 & 61.11 & 49.82 \\
		\rowcolor[HTML]{EFEFEF} 
		Energy (B) & 46.82 & 49.55 & 29.55 & 67.36 & 58.33 & 47.34 \\
		\rowcolor[HTML]{FFFFFF} 
		Duration (C) & 42.47 & 38.63 & 31.02 & 55.41 & 46.68 & 41.75 \\
		\rowcolor[HTML]{EFEFEF} 
		XLS-R & 28.49 & 25.26 & 21.58 & 45.33 & 39.82 & 41.10 \\
		\rowcolor[HTML]{FFFFFF} 
		A + B + C + LFCC & 38.73 & 31.58 & 34.57 & 45.39 & 42.15 & 40.96 \\
		\rowcolor[HTML]{EFEFEF}  
		A + B + C + XLS-R & 29.71 & 22.69 & 20.19 & 40.18 & 36.22 & 34.56 \\
		\bottomrule[1pt]
	\end{tabular}
}
\end{table}

\begin{table}[]
	\centering
	  \caption{ The performance of representative features is evaluated using the detection models trained with the training set of ADD 2023 in terms of the EER (\%). All the models are evaluated on the test set of FG-D task in ADD 2023 and the test set In-the-Wild, respectively. }
        \label{table:features-add}
        \fontsize{6pt}{8}\selectfont
	\resizebox{0.5\textwidth}{!}{
	\begin{tabular}{ccccccc}
		\toprule[1pt]
		\rowcolor[HTML]{FFFFFF} 
		\textbf{Features} & \multicolumn{3}{c}{\cellcolor[HTML]{FFFFFF}\textbf{ADD 2023 FG-D}} & \multicolumn{3}{c}{\cellcolor[HTML]{FFFFFF}\textbf{In-the-Wild}} \\ 
		\cmidrule(r){2-4} \cmidrule(r){5-7}
		\rowcolor[HTML]{FFFFFF} 
		\textbf{} & \textbf{GMM} & \textbf{LCNN} & \textbf{ASSERT} & \textbf{GMM} & \textbf{LCNN} & \textbf{ASSERT} \\ \hline
		\rowcolor[HTML]{EFEFEF} 
		MFCC & 58.36 & 62.13 & 33.78 & 51.80 & 67.62 & 39.79 \\
		\rowcolor[HTML]{FFFFFF} 
		LPS & 60.72 & 57.14 & 42.64 & 65.48 & 50.22 & 24.73 \\
		\rowcolor[HTML]{EFEFEF} 
		LFCC & 63.80 & 65.73 & 36.61 & 50.84 & 52.01 & 45.15 \\
		\rowcolor[HTML]{FFFFFF} 
		IMFCC & 75.44 & 71.85 & 34.62 & 42.66 & 48.94 & 75.47 \\
		\rowcolor[HTML]{EFEFEF} 
		CQCC & 68.29 & 64.17 & 26.81 & 58.54 & 60.95 & 33.35 \\
		\rowcolor[HTML]{FFFFFF} 
		F0 (A) & 62.73 & 72.59 & 39.48 & 71.26 & 63.82 & 47.51 \\
		\rowcolor[HTML]{EFEFEF} 
		Energy (B) & 68.52 & 68.24 & 42.68 & 58.42 & 55.71 & 42.69 \\
		\rowcolor[HTML]{FFFFFF} 
		Duration (C) & 66.18 & 69.15 & 32.60 & 58.65 & 43.66 & 42.93 \\
		\rowcolor[HTML]{EFEFEF} 
		XLS-R & 41.83 & 35.61 & 34.64 & 52.06 & 46.51 & 45.25 \\
		\rowcolor[HTML]{FFFFFF}  
		A + B + C + LFCC & 46.21 & 38.44 & 36.51 & 49.93 & 43.84 & 38.20 \\
		\rowcolor[HTML]{EFEFEF}
		A + B + C + XLS-R & 42.74 & 34.26 & 31.11 & 46.37 & 39.63 & 38.64 \\
		\bottomrule[1pt]
	\end{tabular}
}
\end{table}

For our classifiers, the GMM\footnote{\url{https://github.com/asvspoof-challenge/2021}} model performs binary classification based on a randomly-initialized 512-component GMM, trained with the expectation-maximization algorithm. The LCNN\footnote{\url{https://github.com/asvspoof-challenge/2021}} backend classifier is based on~\cite{2019ASVspoof}, but includes an LSTM layer and average pooling. The ASSERT\footnote{\url{https://github.com/jefflai108/ASSERT}} classifier is based on~\cite{Lai2019ASSERTAW}.

With the aforementioned set of systems, we perform two sets of experiments. In the first set, we train the classifiers on the ASVspoof 2021 DF task training set, and evaluate them on the ASVspoof 2021 DF task test set and the In-the-Wild test set. In the second set, we train the classifiers on the ADD 2023 FG-D task training set, and evaluate them on the ADD 2023 FG-D task test set and the In-the-Wild test set. The results are shown in Tables~\ref{table:features-asv} and~\ref{table:features-add}, respectively.

From the results, it is evident that detection systems perform worse in out-of-domain evaluations compared to in-domain tests, with the EER increasing by 2\%--52\%. It is of considerable interest to note in particular that, while handcrafted features achieve comparable performance during in-domain evaluation, their performance shows greater variability in out-of-domain situations, especially when the test set exhibits greater variance in distribution (the ADD and In-the-Wild data are of different languages), as summarised in Table~\ref{table:features-add}. On the other hand, features from pretrained models, as well as concatenated features, show more consistency in their performance. This suggests that features from pretrained models are more robust to out-of-domain evaluations, and that concatenation of features may be a viable strategy to improve the robustness of detection systems. Notably, the XLS-R features and the concatenated features (A + B + C + XLS-R) as well as LFCC achieve consistent and competitive performance in both in-domain and out-of-domain evaluations.

\begin{table}[]
	\centering
	\caption{ The performance of representative classifiers is evaluated using the detection models trained with the training set of ASVspoof 2021 in terms of the EER (\%). All the models are evaluated on the test set of DF task in ASVspoof 2021 and the test set In-the-Wild, respectively. }
	\label{table:clasifier-asv}
	\fontsize{6pt}{8}\selectfont
	\begin{tabular}{ccccc}
		\toprule[1pt]
		\rowcolor[HTML]{FFFFFF} 
		\textbf{Classifiers} & \multicolumn{2}{c}{\cellcolor[HTML]{FFFFFF}\textbf{ASVspoof 2021 DF}} & \multicolumn{2}{c}{\cellcolor[HTML]{FFFFFF}\textbf{In-the-Wild}} \\
		\cmidrule(r){2-3} \cmidrule(r){4-5} 
		\rowcolor[HTML]{FFFFFF} 
		& \textbf{LFCC} & \textbf{XLS-R} & \textbf{LFCC} & \textbf{XLS-R} \\ \hline
		\rowcolor[HTML]{EFEFEF} 
		GMM & 25.25 & 28.49& 37.49 & 45.33 \\
		\rowcolor[HTML]{FFFFFF} 
		LCNN & 33.74 & 25.26 & 35.14 & 39.82 \\
		\rowcolor[HTML]{EFEFEF} 
		ResNet & 33.42 & 23.83 & 42.17 & 46.35 \\
		\rowcolor[HTML]{FFFFFF} 
		ASSERT & 39.67 & 21.58 & 48.82 & 41.10 \\
		\rowcolor[HTML]{EFEFEF} 
		Res2Net & 35.18 & 19.47 & 39.18 & 36.62 \\
		\rowcolor[HTML]{FFFFFF} 
		AFN & 38.58 & 14.15 & 30.67 & 42.46 \\
		\rowcolor[HTML]{EFEFEF} 
		GRU &56.06  & 56.15 & 51.68 & 49.56 \\
		\rowcolor[HTML]{FFFFFF} 
		GAT & 47.14 & 14.91 & 33.46 & 44.31  \\
		\bottomrule[1pt]
	\end{tabular}
\end{table}

\begin{table}[]
	\centering
	\caption{ The performance of representative classifiers is evaluated using the detection models trained with the training set of ADD 2023 in terms of the EER (\%). All the models are evaluated on the test set of FG-D task in ADD 2023 and the test set In-the-Wild, respectively. }
	\label{table:clasifier-add}
	\fontsize{6pt}{8}\selectfont
	\begin{tabular}{ccccc}
		\toprule[1pt]
		\rowcolor[HTML]{FFFFFF} 
		\textbf{Classifiers} & \multicolumn{2}{c}{\cellcolor[HTML]{FFFFFF}\textbf{ADD 2023 FG-D}} & \multicolumn{2}{c}{\cellcolor[HTML]{FFFFFF}\textbf{In-the-Wild}} \\
		\cmidrule(r){2-3} \cmidrule(r){4-5} 
		\rowcolor[HTML]{FFFFFF} 
		& \textbf{LFCC} & \textbf{XLS-R} & \textbf{LFCC} & \textbf{XLS-R} \\ \hline
		\rowcolor[HTML]{EFEFEF} 
		GMM & 63.80 & 41.83 & 50.84 & 52.06 \\
		\rowcolor[HTML]{FFFFFF} 
		LCNN & 65.73 & 35.61 & 52.01 & 46.51 \\
		\rowcolor[HTML]{EFEFEF} 
		ResNet & 52.13 & 36.85 & 49.57 & 43.22 \\
		\rowcolor[HTML]{FFFFFF} 
		ASSERT & 36.61 & 34.64 & 45.15 & 45.25 \\
		\rowcolor[HTML]{EFEFEF} 
		Res2Net & 48.62 & 36.91 & 48.63 & 45.42 \\
		\rowcolor[HTML]{FFFFFF} 
		AFN & 38.36 & 37.50 & 42.54 & 45.22 \\
		\rowcolor[HTML]{EFEFEF} 
		GRU & 59.25 & 52.70 & 65.63 & 55.77 \\
		\rowcolor[HTML]{FFFFFF} 
		GAT & 42.13 & 36.97 & 44.98 & 47.76  \\
		\bottomrule[1pt]
	\end{tabular}
\end{table}

\subsection{Evaluation of Classifiers}

Following the previous evaluations of discriminative features, we also evaluate the performances of different back-end classifiers and end-to-end models. We choose the GMM, LCNN, ResNet, ASSERT, Res2Net, AFN and GAT classifiers due to their popularity and effectiveness in fake audio detection tasks. Owing to the effectiveness of the XLS-R features and LFCC, we evaluate the classifiers on these two features.

The GMM, LCNN and ASSERT classifiers are configured as described in the previous section. The ResNet classifier is a ResNet-34 model. The Res2Net\footnote{\url{https://github.com/lixucuhk/ASV-anti-spoofing-with-Res2Net}}, AFN\footnote{\url{https://github.com/jefflai108/Attentive-Filtering-Network}}, GRU\footnote{\url{https://pytorch.org/docs/stable/generated/torch.nn.GRU.html}} and GAT\footnote{\url{https://github.com/TakHemlata/SSL_Anti-spoofing}} classifiers are based on the open-source implementations. In training, we use the Adam optimizer with a learning rate of \(5 \times 10^{-5}\); the models are trained for 200 epochs with a batch size of 32. The results are shown in Tables~\ref{table:clasifier-asv} and~\ref{table:clasifier-add}.

As seen in Table~\ref{table:clasifier-asv}, the AFN and GAT classifiers perform very well under in-distribution conditions, with the lowest EER of 14\%--15\%; the classifiers do not show significant variance in their performance under out-of-distribution conditions. This shows that deep convolutional models with pretrained model features are effective when the testing data is in-distribution. However, the performance of the classifiers degrades significantly when there is a discrepancy between the distributions of the training and testing data. This is especially evident in the performance of the GAT classifier, which shows a 10\% increase in EER when evaluated on the In-the-Wild test set. This suggests that the classifiers are not as robust to out-of-distribution evaluations.

This is further evidenced by the results in Table~\ref{table:clasifier-add}, where the classifiers are evaluated on the ADD 2023 FG-D task test set and the In-the-Wild test set. The classifiers show a significant increase in EER when evaluated on the In-the-Wild test set, which is of a different language from the training dataset. This suggests that the classifiers are not robust to out-of-distribution evaluations, and that the performance of the classifiers degrades significantly when the testing data is out-of-distribution. 

\begin{table}[]
	\centering
	\caption{ The performance of end-to-end models is evaluated using the detection models in terms of the EER (\%). All models are evaluated on the test set of ASVspoof 2021 DF task, ADD 2023 FG-D task, and In-the-Wild.}
        \label{table:end-to-end}
	\fontsize{6pt}{8}\selectfont
	\begin{tabular}{ccccc}
		\toprule[1pt]
		\multicolumn{1}{c}{} & \multicolumn{2}{c}{\textbf{Trained on ASVspoof 2021}} & \multicolumn{2}{c}{\textbf{Trained on ADD 2023}} \\
		\cmidrule(r){2-3} \cmidrule(r){4-5} 
		\multicolumn{1}{c}{\multirow{-2}{*}{\textbf{Models}}} &  \multicolumn{1}{l}{\textbf{ASVspoof 2021 DF}} & \multicolumn{1}{l}{\textbf{In-the-Wild}} & \textbf{ADD 2023 FG-D} & \textbf{In-the-Wild} \\ \hline
		\rowcolor[HTML]{EFEFEF} 
		RawNet2 & 24.32 & 36.74 & 54.51 & 40.35 \\
		AASIST & 19.77 & 34.81 & 48.66 & 37.63 \\ \bottomrule[1pt]
	\end{tabular}
\end{table}

\subsection{End-to-End Models}

We evaluate the performances of different end-to-end models like RawNet2, AASIST end-to-end models due to their popularity and effectiveness in fake audio detection tasks. RawNet2\footnote{\url{https://github.com/asvspoof-challenge/2021/tree/main/LA/Baseline-RawNet2}}, 
ASIST\footnote{\url{https://github.com/clovaai/aasist}} end-to-end models are based on the open-source codes. 
In training, we use the Adam optimizer with a learning rate of \(5 \times 10^{-5}\); the models are trained for 200 epochs with a batch size of 32. 

Table~\ref{table:end-to-end} shows the performance of the end-to-end models, which are trained on the ASVspoof 2021 DF task training set as well as the ADD 2023 FG-D task training set, and evaluated on the ASVspoof 2021 DF task test set, the ADD 2023 FG-D task test set, and the In-the-Wild test set. The results obtained from models trained on the ASVspoof dataset show that the end-to-end models perform well under in-distribution conditions, with the lowest EER of 19\%--25\%. However, the performance of the end-to-end models degrades when evaluated on the In-the-Wild test set, with the EER increasing by 10\%--15\%. 

On the other hand, the models trained on the ADD 2023 dataset perform better on In-the-Wild, due to the difference between the ADD 2023 training and testing sets in terms of audio quality and perturbations, possibly resulting in greater distribution differences. 
Overall, these results suggest that the addressing of this issue is a promising direction for future research.

\section{Future Directions}

Although some significant progress on audio deepfake detection have been made over the last decade, there still exist some limitations which should be addressed in future work. Some potential research directions are summarized as follows.

\textbf{Collecting audio datasets in the wild:} Most of the audio deepfake detection datasets are not collected in the wild, which do not quite match with the real utterances recorded or generated in realistic conditions. The real conditions of the utterances may be even worse and vary more greatly than the simulated conditions. In order to assess audio deepfake detection methods in practical applications, the utterances with a variety of channels or conditions should be collected through realistic environment conditions, such as social media platforms, Internet or telephone channels.

\textbf{Designing large-scale multilingual datasets}: Previous datasets are mainly single language based and most of them are English deepfake audio datasets and few of them are other language datasets, e.g. Chinese or Japanese. It may make the detection methods language dependent. But it is necessary to build language independent detection systems in realistic applications. In order to make fake detection systems more robust for other languages, we need to evaluate the performance of fake detection models in the cross language scenario and for code-switching between different languages. Therefore, it is important to design and develop large-scale multilingual datasets on audio deepfake detection in the wild.

\textbf{Improving generalization ability and robustness of detection models}: Although previous studies have made a lot of attempts on audio deepfake detection and achieved promising performance, the generalization and robustness of the detection models are still poor. The performance of the top-performing models in the ASVspoof and ADD competitions are very high but it will degrade significantly when evaluated on the mismatching dataset containing unseen fake attacks or unseen acoustic conditions or unseen language. Some researchers have use effective loss function or continual learning methods to address this problem but there is still much room for improvement. 

\textbf{Dealing with rapid development of deepfake technologies}: Deepfake technologies are developing rapidly and the generated audio becoming increasingly realistic, it is hard to detect correctly. It brings new challenges to current existing detection methods. In order to alleviate this issue, deepfake audio generation and detection task are viewed as a rivalry game for participants in the ADD 2022 and 2023 competitions. Participants in the generation task aim to generate deepfake samples to fool the detection model while participants in the detection task try to detect all fake utterances as much as possible. Despite partly improving the anti-attack ability of the detection model via fake game, the methods of game are simple and lack intelligence. Therefore, we should propose more effective detection approaches to copy with the new unseen deepfake aduio technologies.

\textbf{Improving the interpretability of detection results}: 
Most of existing researches foucs on distinguishing the fake audio from the bona fide one. 
However, there is also an interest in surpassing the constraints of binary real/fake classification, and actually localizing the manipulated intervals in a partially fake speech as well as pinpointing the source responsible for generating any fake audio. In addition, interpretability of detection results is needed to provide in real applications, e.g. audio forensics and attribution. It is nontrivial to know why the utterance is fake and find the tools or deepfake methods of generating the fake audio. In addition, it is also important to know what manipulation technologies are employed and even intention of the manipulation. It is particularly critical for audio forensics. Despite ADD 2023 challenge including manipulation region location and deepfake algorithm recognition sub-challenges, the studies are still in infancy.

\textbf{Exploring more reasonable evaluation metrics}: EER is widely employed as the evaluation metric in previous work, such as ASVspoof and ADD competitions. However, we need to assess whether the EER is reasonable for the audio deepfake detection model or not in the future. We should consider human detection capabilities, as well as the differences between humans and machines for detecting deepfake audio.

\section{Conclusions}

The deepfake technologies pose a serious threat to social security and political economy if someone misuses them for malicious purposes. 
Therefore, it is indispensable to detect deepfake audio. To make audio deepfake detection useful in practice, we need to propose robust and general algorithms with valid
and reliable samples in order to make the detection of deepfake audio applicable to real situations. Accordingly, in this survey, we review the current research on audio deepfake detection. We further compare the performance of existing state-of-the-art methods, analyze the potential, and highlight the outstanding issues for future research. Audio deepfake detection has recently become an active research area; accordingly, we hope this survey can help researchers, as a starting point, to review the developments in the state-of-the-art and identify possible directions for their future research.


%


\ifCLASSOPTIONcaptionsoff
  \newpage
\fi



%



\bibliographystyle{IEEEtran}
\bibliography{Mybib}

%

\begin{IEEEbiography}[{\includegraphics[width=1in,height=1.25in,clip,keepaspectratio]{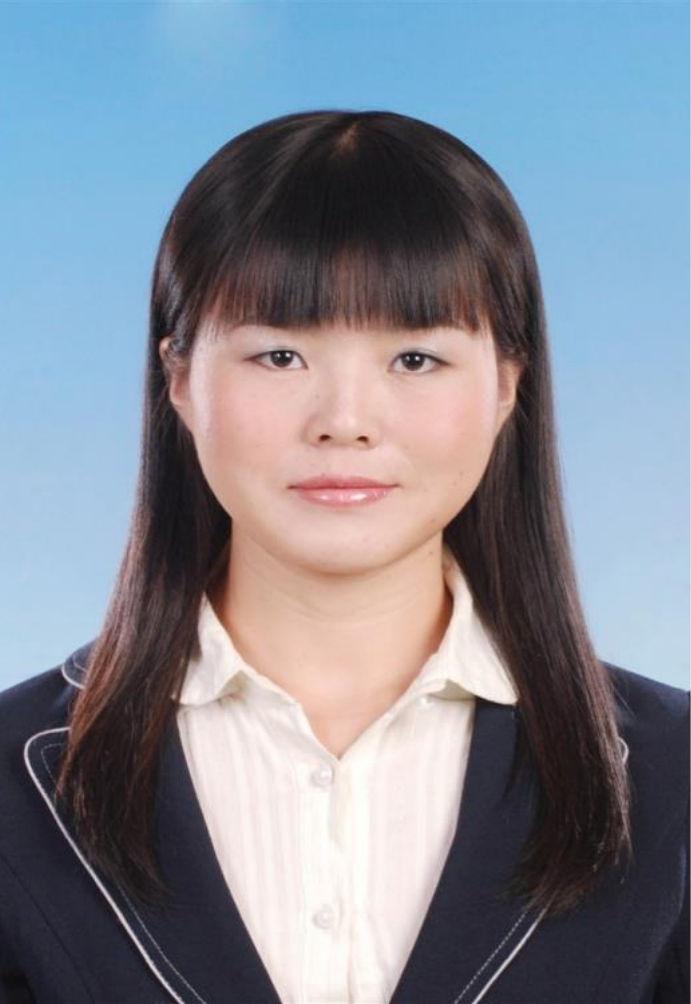}}]{Jiangyan Yi}
received the Ph.D. degree from the University of Chinese Academy of Sciences in 2018, and the M.A. degree from the Graduate School of Chinese Academy of Social Sciences in 2010. She was a Senior R\&D Engineer with Alibaba Group during 2011 to 2014. She is currently an Associate Professor with the State Key Laboratory of Multimodal Artificial Intelligence Systems, Institute of Automation, Chinese Academy of Sciences. Her current research interests include speech signal processing, speech recognition and synthesis, fake audio detection, audio forensics and transfer learning.
\end{IEEEbiography}

\begin{IEEEbiography}[{\includegraphics[width=1in,height=1.25in,clip,keepaspectratio]{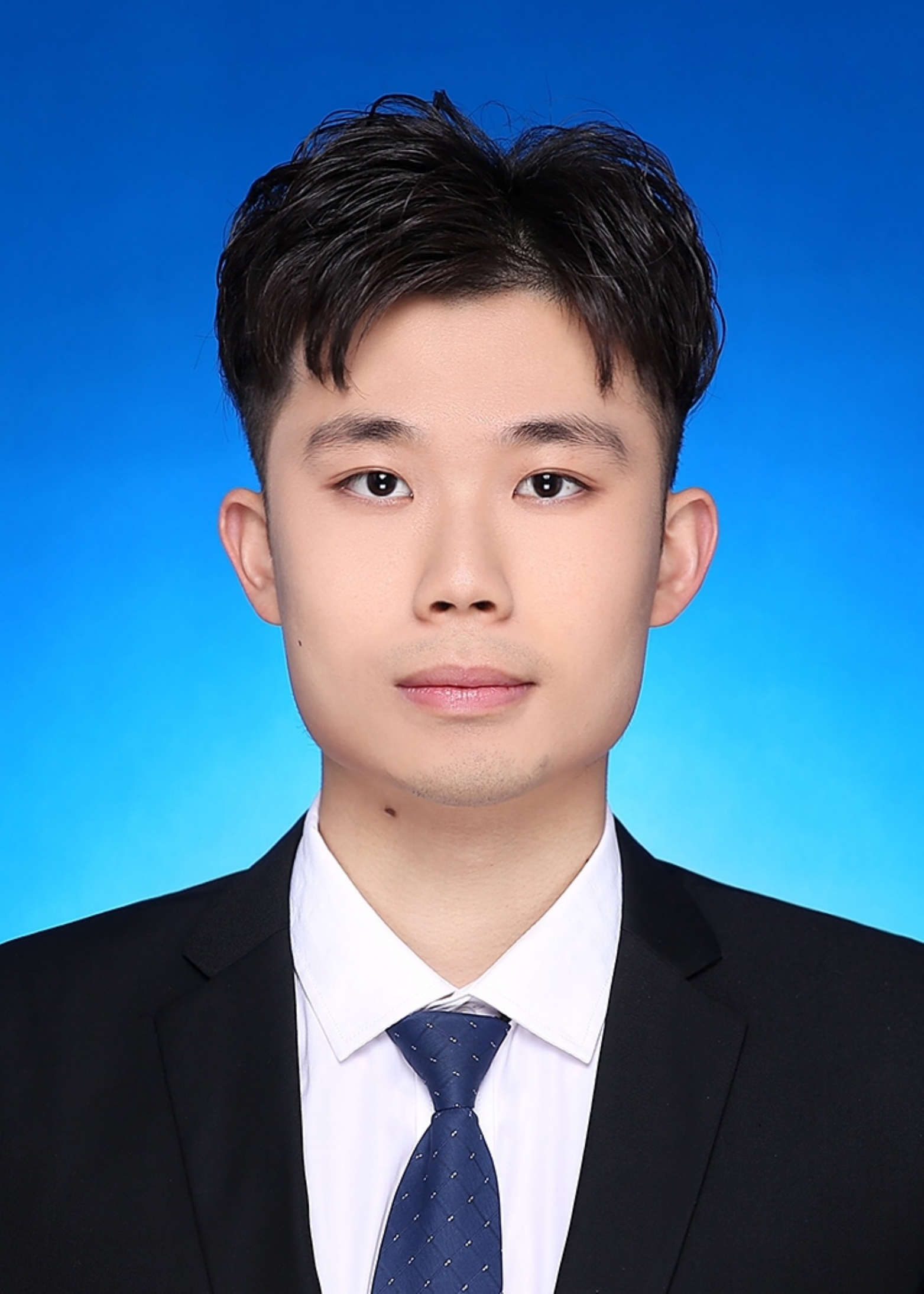}}]{Chenglong Wang}
 received the B.S. degree from Hefei University of Technology, Anhui, China, in 2018. He is currently working toward the Ph.D. degree with the University of Science and Technology of China, Anhui, China. His current research interests include fake audio detection, speaker verification and identification.
\end{IEEEbiography}

\begin{IEEEbiography}[{\includegraphics[width=1in,height=1.25in,clip,keepaspectratio]{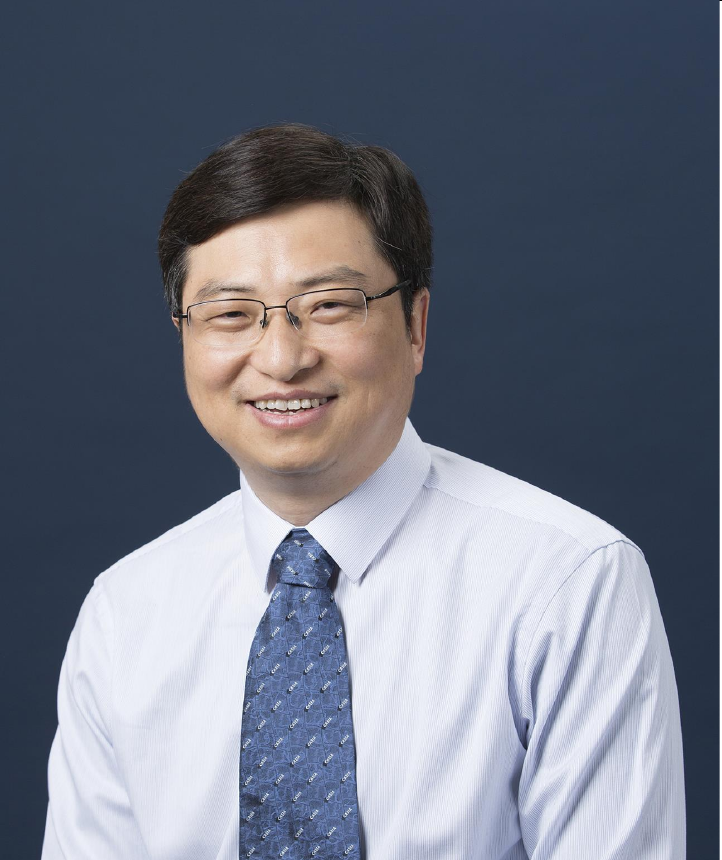}}]{Jianhua Tao}
 received his Ph.D. degree from Tsinghua University, Beijing, China, in 2001, and the M.S. degree from Nanjing University, Nanjing, China, in 1996. He is currently a Professor with the Department of Automation, Tsinghua University. He has authored or coauthored more than eighty papers on major journals and proceedings. His current research interests include speech signal processing, speech recognition and synthesis, human computer interaction, multimedia information processing, and pattern recognition.
\end{IEEEbiography}

\begin{IEEEbiography}[{\includegraphics[width=1in,height=1.25in,clip,keepaspectratio]{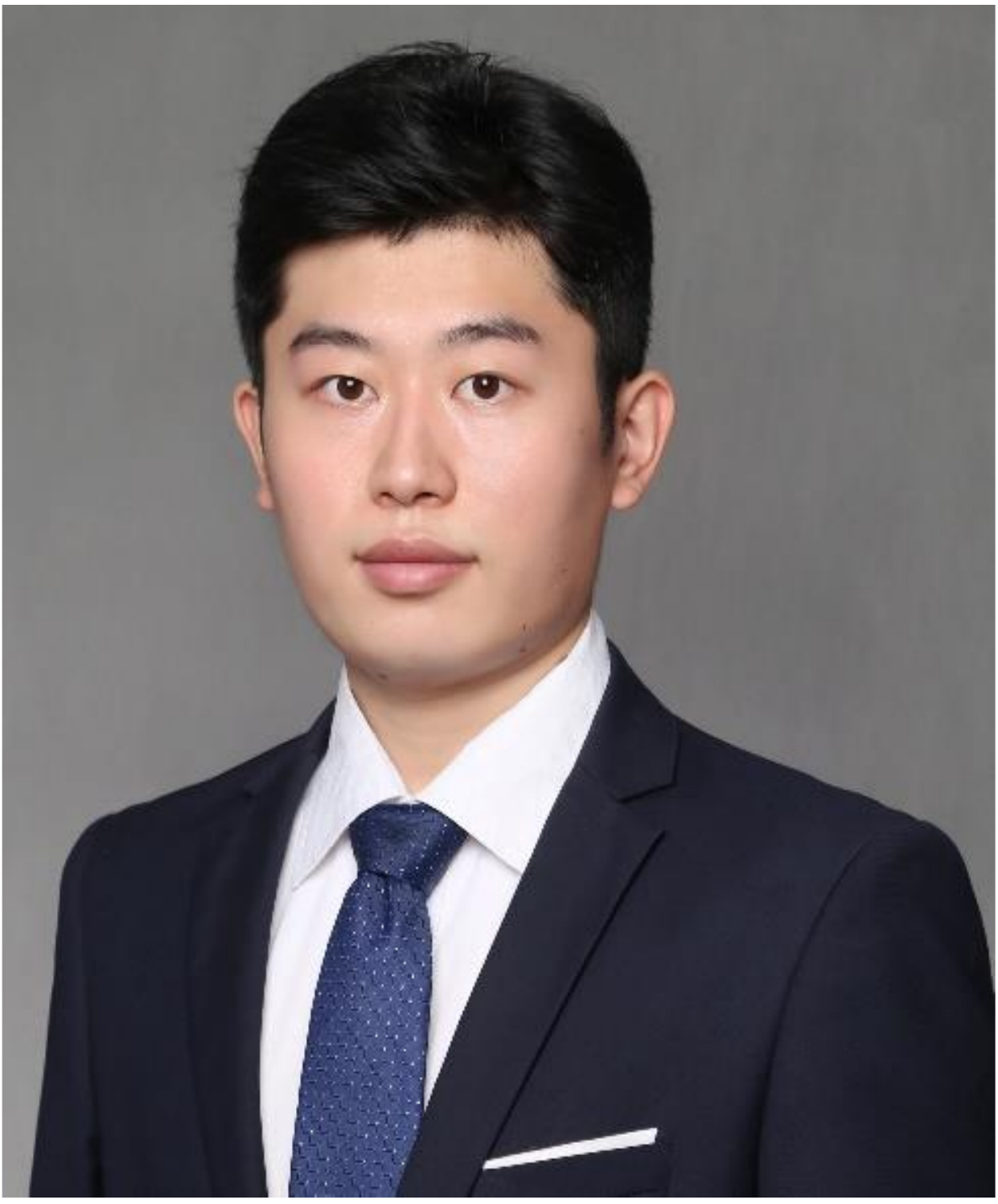}}]{Xiaohui Zhang}
 received the B.S.degree from Beijing Jiaotong University, Beijing, China, in 2021. He is currently working toward a master’s degree at Beijing Jiaotong University, Beijing, China. His current research interests include fake audio detection and continual learning.
\end{IEEEbiography}

\begin{IEEEbiography}[{\includegraphics[width=1in,height=1.25in,clip,keepaspectratio]{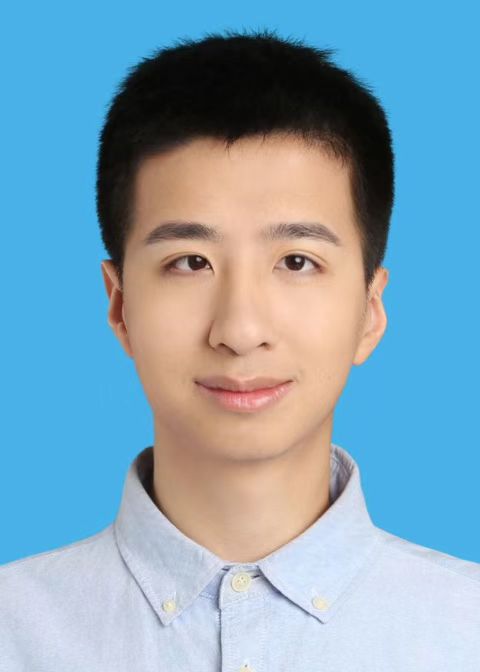}}]{Chu Yuan Zhang}
received his B.A. degree in linguistics at the University of California, Los Angeles (UCLA), in 2021. He is currently pursuing a M.E. degree at the Institute of Automation, Chinese Academy of Sciences and the International College at the University of Chinese Academy of Sciences. His current research interests include TTS and machine learning.
\end{IEEEbiography}

\begin{IEEEbiography}[{\includegraphics[width=1in,height=1.25in,clip,keepaspectratio]{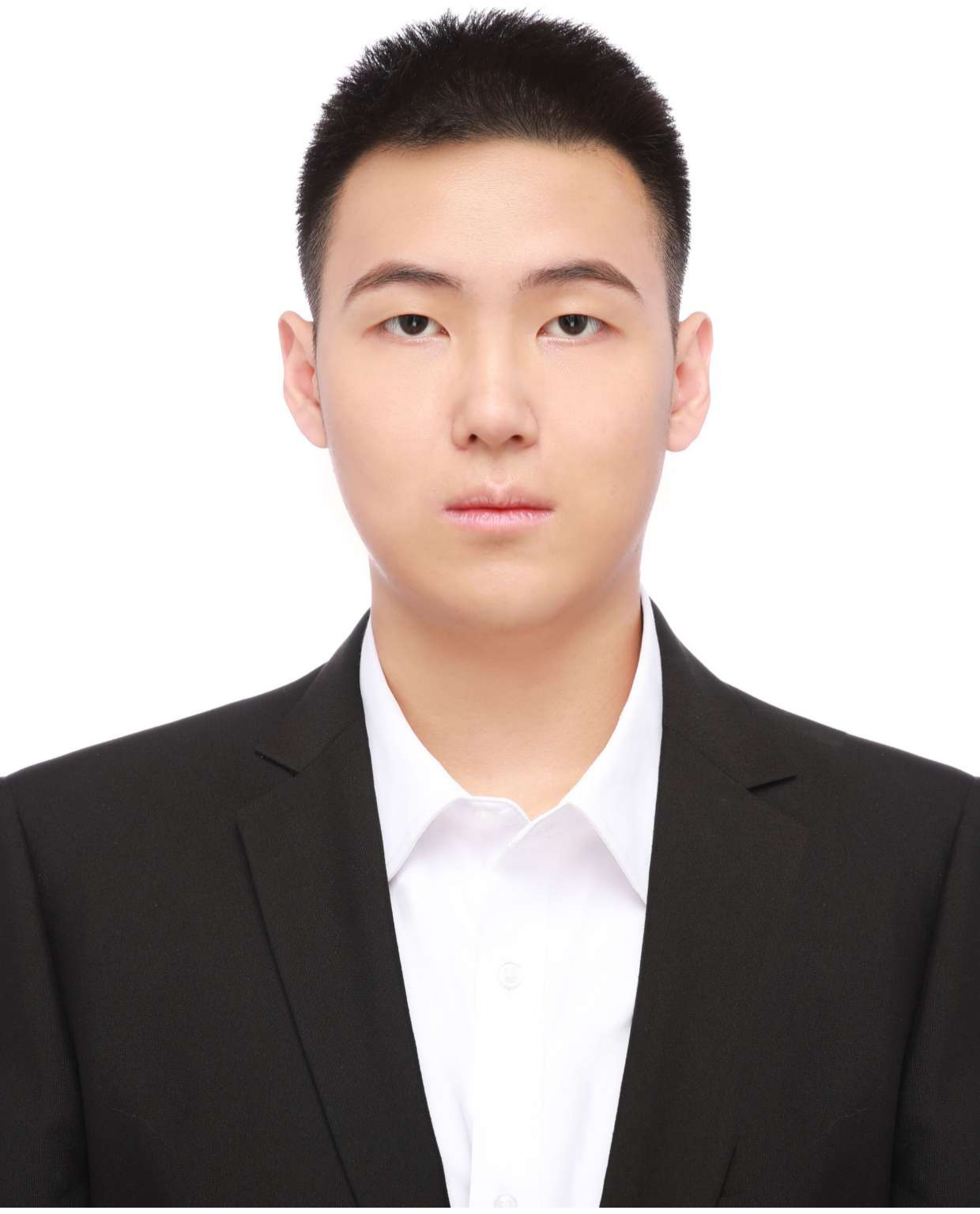}}]{Yan Zhao}
obtained a B.S degree from Inner Mongolia University, in 2021. He is currently pursuing a M.S. degree at Hebei University of Technology. His research interests include speech adversarial sample attacks and fake audio detection.
\end{IEEEbiography}






\end{document}